\documentclass[11pt]{article}
\pdfoutput=1 
\usepackage{jheppub}
\usepackage[utf8]{inputenc}

\usepackage{graphicx,ulem}
\usepackage{caption,epigraph}
\usepackage{subcaption}
\usepackage{hyperref}
\usepackage[T1]{fontenc}
\usepackage[export]{adjustbox}
\usepackage[table]{xcolor}
\usepackage{amssymb}
\usepackage{pifont}

\newcommand{\be}{\begin{equation}}
\newcommand{\ee}{\end{equation}}
\newcommand{\bea}{\begin{eqnarray}}
\newcommand{\eea}{\end{eqnarray}}

\DeclareUnicodeCharacter{2212}{-}
\usepackage{multirow, graphicx,amssymb,url,mathrsfs,amsmath}
\usepackage{amsxtra,amstext,latexsym,dsfont,amsfonts}
\usepackage{color,eucal}
\usepackage {verbatim}
\usepackage{float}
\usepackage{colortbl}
\usepackage{multirow}
\usepackage{tikz}
\usetikzlibrary{tikzmark}
\usetikzlibrary{arrows}
\usepackage{tabularx, array}
\newcolumntype{L}[1]{>{\raggedright\arraybackslash}p{#1}}
\newcolumntype{C}[1]{>{\centering\arraybackslash}p{#1}}
\newcolumntype{R}[1]{>{\raggedleft\arraybackslash}p{#1}}
\makeatother
\makeatletter
\newcommand{\myBig}{\bBigg@{1.75}}
\makeatother

\title{\Large Breaking rotations without violating the KSS viscosity bound}

\author[a,b,c]{Matteo Baggioli,}
\author[d,e]{Sera Cremonini,}
\author[d]{Laura Early,}
\author[f,g,h,i]{Li Li,}
\author[f,g]{Hao-Tian Sun}
\affiliation[a]{School of Physics and Astronomy, Shanghai Jiao Tong University, Shanghai 200240, China}
\affiliation[b]{Wilczek Quantum Center, School of Physics and Astronomy, Shanghai Jiao Tong University, Shanghai 200240, China}
\affiliation[c]{Shanghai Research Center for Quantum Sciences, Shanghai 201315, China}
\affiliation[d]{Department of Physics, Lehigh University, Bethlehem, PA, 18015, USA}
\affiliation[e]{
Kavli Institute of Theoretical Physics, University of California Santa Barbara, Santa Barbara,CA, 93106, USA}
\affiliation[f]{CAS Key Laboratory of Theoretical Physics, Institute of Theoretical Physics,
Chinese Academy of Sciences, Beijing 100190, China}
\affiliation[g]{School of Physical Sciences, University of Chinese Academy of Sciences, Beijing 100049, China}
\affiliation[h]{School of Fundamental Physics and Mathematical Sciences, Hangzhou Institute for Advanced Study, University of Chinese Academy of Sciences, Hangzhou 310024, China}
\affiliation[i]{Peng Huanwu Collaborative Center for Research and Education, Beihang University, Beijing 100191, China.}

\emailAdd{b.matteo@sjtu.edu.cn}
\emailAdd{cremonini@lehigh.edu}
\emailAdd{lte219@lehigh.edu}
\emailAdd{liliphy@itp.ac.cn}
\emailAdd{sunhaotian@itp.ac.cn}

\abstract{We revisit the computation of the shear viscosity to entropy ratio in a holographic p-wave superfluid model, focusing on the role of rotational symmetry breaking. We study the interplay between explicit and spontaneous symmetry breaking and derive a simple horizon formula for $\eta/s$, which is valid also in the presence of explicit breaking of rotations and is in perfect agreement with the numerical data. 
We observe that a source which explicitly breaks rotational invariance suppresses the value of $\eta/s$ in the broken phase, competing against the effects of spontaneous symmetry breaking.
However, $\eta/s$ always reaches a constant value in the limit of zero temperature, which is never smaller than the Kovtun-Son-Starinets (KSS) bound, $1/4\pi$. This behavior appears to be in contrast with previous holographic anisotropic models which found a power-law vanishing of $\eta/s$ at small temperature. 
This difference is shown to arise from the properties of the near-horizon geometry in the extremal limit. Thus, our construction shows that the breaking of rotations itself does not necessarily imply a violation of the KSS bound.
}

\begin{document} 

\maketitle
\section{Introduction}

One of the most quantitative early applications of holography to 
strongly correlated systems has been
the realization that 
the shear viscosity $\eta$ to entropy density $s$ obeys a simple result, 
\be
\label{universal}
\frac{\eta}{s} = \frac{\hbar}{4\pi k_B} \, ,
\ee
which is universal in a large class of theories \cite{Policastro:2001yc,Buchel:2003tz}.
Despite the compelling proposal \cite{Kovtun:2003wp,Kovtun:2004de} that this simple ratio could be a fundamental lower bound in nature, it has been well understood that the so-called KSS bound can be violated in a number of ways (see \cite{Cremonini:2011iq} for a review).
The first violations \cite{PhysRevLett.99.021602,Cherman:2007fj,Dobado:2007tm} (which likely lack a well-defined UV relativistic completion \cite{PhysRevLett.100.029101}) were obtained by considering non-relativistic systems with a large number of species.\footnote{Notice, however, that for non-relativistic classical liquids, a different bound on the kinematic viscosity has been recently proposed and verified \cite{doi:10.1126/sciadv.aba3747,PhysRevB.103.014311}.}
On the other hand, without relaxing Poincar\'e symmetry, certain higher derivative corrections to the low-energy Einstein action can push $\eta/s$ below its universal value, in a controlled way \cite{Buchel:2004di,Buchel:2008vz,Kats:2007mq,Myers:2008yi} (\emph{i.e.}, the corrections are perturbatively small and a minimum value different from zero still exists). Indeed, such higher derivative operators are well motivated by top-down string theory constructions, 
and encode $1/N$ effects in the dual gauge theory, with $N$ the number of colors. Within these models, it has been often argued that causality 
and stability in the UV of the theory are the key features behind the existence of a finite, yet non universal, minimum \cite{Brigante:2007nu,Brigante:2008gz}.  
Nevertheless, see \cite{Buchel:2010wf} for an early counterexample to the statement that UV properties are necessarily linked to a possible lower bound on $\eta/s$.

Violations can also be realized within Einstein gravity, without having to invoke higher derivative operators, by working with setups that break spacetime symmetries (translations and/or rotations). In these scenarios, the deviations are more drastic since the $\eta/s$ ratio generally vanishes at zero temperature following a power-law $(T/\gamma)^{\#}$ where $\#>0$ and $\gamma$ is the scale parameterizing the strength of anisotropy or translational symmetry breaking (\underline{e.g.}, \cite{Ling:2016ien}). In the case of translations, the physical interpretation of this phenomenon remains obscure as it is made complicated by the fact that either momentum is not conserved \cite{Hartnoll:2016tri,Alberte:2016xja,Burikham:2016roo} or the dual field theory is no longer in a liquid phase \cite{Baggioli:2020ljz,Baggioli:2022aft,RevModPhys.95.011001}.\footnote{However, the $\eta/s$ ratio can be still understood as the rate of entropy production due to strain \cite{Hartnoll:2016tri}. Also notice that, although the $\eta/s$ ratio violates the KSS bound, the momentum diffusivity does not \cite{Baggioli:2020ljz}.}

Bottom-up models that describe anisotropic phases have also led to large deviations from (\ref{universal}), and a temperature-dependent behavior that is sensitive to the particular details of the model. The first holographic model to observe a violation of the $\eta/s$ bound due to anisotropy is the non-commutative plasma of Ref. \cite{Landsteiner:2007bd}. Following this observation, the violation of the KSS bound has been observed in many holographic anisotropic models. A popular subclass of such constructions relies on the introduction of a bulk axion field (or of a higher-form generalization of the latter \cite{Liu:2016njg}) whose profile selects a specific spatial direction in the boundary field theory \cite{Giataganas:2017koz,Donos:2016zpf,Ge:2015owa,Giataganas:2013lga,Chakraborty:2017msh,Mateos:2011ix,Rebhan:2011vd,Jain:2014vka,Jain:2015txa,Azeyanagi:2009pr}. A common alternative consists in breaking rotational invariance using an external magnetic field \cite{Rath:2020beo,Gursoy:2020kjd,Finazzo:2016mhm,Critelli:2014kra,Ammon:2012qs}. A third possibility is to break  rotations spontaneously in a system undergoing a p-wave superfluid instability \cite{Natsuume:2010ky,Basu:2011tt,Erdmenger:2011tj,Bhattacharyya:2014wfa,Oh:2012zu}. Moreover, the KSS bound can be violated in holographic Weyl-semimetals \cite{Landsteiner:2016stv}, in holographic models for tilted Dirac materials \cite{Moradpouri:2022zwa} as well as in anisotropic top-down models (\underline{e.g.}, \cite{Polchinski:2012nh,Penin:2017lqt}). Interestingly, violations of the KSS bound have
also been discussed in pure condensed matter models \cite{Samanta:2016pic,kim2021hydrodynamic,gochan2019viscosity,Link:2017ora}. Finally, violations of the KSS bound have been reported in out-of equilibrium holographic systems \cite{Wondrak:2020tzt,Baggioli:2021tzr} where, nevertheless, the definition of the shear viscosity becomes less obvious.

Within the large class of holographic anisotropic models, a sharp distinction can be made. In particular, anisotropy could be either spontaneous (\underline{e.g.}, ferromagnetic materials) or explicit (\underline{e.g.}, materials under an external magnetic field). 
From a technical perspective, this difference depends on whether the rotational symmetry is broken by the vacuum expectation value of a certain operator, or by the external source associated with it. Both possibilities can be realized holographically, and leave sharply different imprints on $\eta/s$. Indeed, even though both scenarios lead to a deviation from \eqref{universal}, only the latter has been shown to induce a violation of the KSS bound,  $\eta/s<1/4\pi$, with certain models exhibiting a decrease of $\eta/s$ towards zero temperature (see for example \cite{Jain:2015txa,Finazzo:2016mhm}). In the case of spontaneous anisotropy 
examined in 
\cite{Erdmenger:2011tj}, on the other hand,
the $\eta/s$ ratio is larger than the ``universal'' value in \eqref{universal} and grows towards small temperatures.  
We shall discuss these differences in more 
detail below. Finally, the interplay of rotational and translational symmetry breaking (\emph{e.g.}, ordinary crystals) could also play an important role in this discussion. For simplicity, in this work, we will disentangle the two effects by considering holographic models in which translations are preserved.

A second, and equally important, issue relevant to the physics associated with the shear viscosity is its temperature dependence, and in particular the existence of a minimum in $\eta/s$ as a function of temperature. In classical liquids, the presence of a minimum is expected on general grounds \cite{doi:10.1126/sciadv.aba3747}. In particular, using standard kinetic theory, valid for dilute gases, the viscosity is given by $\eta \sim \rho_m v_p l$ where $\rho_m$ is the density, $v_p$ the average particle velocity, and $l$ the mean free path. Since the velocity increases with temperature, $v_p \propto \sqrt{T}$, the viscosity increases as well. On the contrary, in the liquid regime, the viscosity emerges from thermally activated jumps (and not from thermal collisions) and it increases towards lower temperatures
as $\eta \sim \exp \left(U/T\right)$, where $U$ is the activation energy. This brief argument already indicates the existence of a minimum in the viscosity as a function of temperature which is indeed observed in all classical liquids \cite{doi:10.1126/sciadv.aba3747}, strongly coupled plasmas \cite{huang2023revealing,PhysRevResearch.4.033064} and ultracold Fermi gases \cite{Cao:2011fh}. The same minimum is expected to be present also in the quark gluon plasma \cite{PhysRevD.76.101701} (see \cite{Adams:2012th} for an overview). Back to the holographic phenomenology, it is usually quite challenging to obtain a non-monotonic behavior of the $\eta/s$ ratio. Neverthless, using a running dilaton bulk field \cite{Cremonini:2012ny} or constructing more complex gravitational solutions interpolating between different scale invariant geometries \cite{Cremonini:2011ej}, it is possible to achieve a minimum of $\eta/s$ as a function of temperature, reminiscent of classical liquids.

In this paper we revisit the question of the behavior of $\eta/s$ in holographic systems that are anisotropic, while preserving the translation symmetry, building on previous work in the literature in a number of ways. We will work with gravitational models in which rotational symmetry can be broken both explicitly and spontaneously.
Our interest is two-fold.
We want to understand the role played by 
different mechanisms of rotational symmetry breaking, and how their interplay
controls the structure of $\eta/s$ and its deviation from the universal value (\ref{universal}).  
In addition, we want to ask whether the 
competition between explicit and spontaneous symmetry breaking
is in fact universal, and whether it could in principle generate a minimum for $\eta/s$ as a function of temperature.
If so, it would provide insights into the mechanisms behind a possible fundamental lower bound on $\eta/s$.

In order to answer these questions, we consider a (five-dimensional) holographic model for p-wave superfluidity, in which a vector condensate breaks simultaneously a U(1) symmetry together with the rotational group SO$(3) \rightarrow$\,SO$(2)$ \cite{Gubser:2008wv} (see \cite{Cai:2015cya} for a review of holographic p-wave superfluids). Unfortunately, in this model it is not possible to break solely the rotational symmetry, which is always ``slaved'' to the U(1). Nevertheless, gaining intuition from the holographic s-wave superfluid case \cite{Natsuume:2010ky}, we do not expect the breaking of the U(1) symmetry to violate the KSS bound. 
In anisotropic fluids, the shear viscosity generalizes to a rank-$4$ tensor, the viscosity tensor, which can be defined as
\begin{align}
    & \eta^{ijkl}\,=\,-\lim_{\omega \rightarrow 0} \,\frac{1}{\omega} \,\mathrm{Im} \,\mathcal{G}^{ijkl}_R\left(\omega,\Vec{0}\right)\,,
\end{align}
where $\mathcal{G}^{ijkl}_R\left(\omega,\Vec{0}\right)$ is the retarded Green's function for the stress tensor operator $T^{ij}$ evaluated at zero wave-vector $\Vec{k}=\Vec{0}$ and finite frequency $\omega$. Because of the specific symmetry breaking pattern, SO$(3) \rightarrow$\,SO$(2)$, the viscosity tensor contains only two independent coefficients, $\eta^{xyxy}$ and $\eta^{yzyz}$. Identifying the anisotropic direction with the $x$ coordinate and following the standard notation in the literature, we denote the two coefficients by $\eta_\parallel\equiv \eta^{yzyz}$ and $\eta_\perp\equiv \eta^{xyxy}$, as they represent, respectively, the viscous friction in the directions 
parallel and perpendicular to the anisotropy. Anisotropic viscosities are widely studied in the context of liquid crystals and nematic liquids \cite{refId0,doi:10.1146/annurev.fl.10.010178.001213,doi:10.1080/00268948108076134}, in which they are usually parameterized using the Miesowicz coefficients \cite{MIESOWICZ1946,doi:10.1063/1.465570}, which turn out to be simple combinations of $\eta_\parallel$ and $\eta_\perp$.

The two viscosities described above have been computed numerically  in holographic p-wave superfluids in \cite{Erdmenger:2010xm,Erdmenger:2011tj,Erdmenger:2012zu}. 
Note that $\eta_\parallel$, which parameterizes the viscosity in the SO$(2)$ invariant $yz$ plane, corresponds to a tensor mode and trivially saturates the KSS bound, Eq.\eqref{universal}. On the contrary, the other viscosity $\eta_\perp$ is strongly affected by the anisotropy and does not obey Eq.\eqref{universal}. Below the critical temperature $T_c$, where the isotropy is lost, $\eta_\perp/s$ is larger than $1/4\pi$ and grows towards zero temperature. 
Moreover, the deviation from $1/4\pi$  is increased by making the backreaction of the SU(2) vector field in the bulk larger. 
As we already mentioned, the  behavior seen in \cite{Erdmenger:2010xm,Erdmenger:2011tj,Erdmenger:2012zu}
is strikingly different from other anisotropic holographic models, in which isotropy is broken explicitly by an external source and the value of $\eta_\perp/s$ violates the KSS bound (\emph{e.g.}, \cite{Jain:2015txa,Finazzo:2016mhm}). In particular, in the latter class of models, $\eta_\perp/s$ becomes smaller than $1/4\pi$ below $T_c$ and vanishes as a power-law towards $T\rightarrow 0$.

To explore this conundrum and understand the origin of this difference, 
in this paper we have modified the original holographic p-wave superfluid model by adding an external source for the vector operator which forms the spontaneous condensate. This gives us a concrete way to study the interplay between the spontaneous and explicit breaking of rotations, and inspect how the behavior of the condensate is imprinted on that 
of $\eta/s$. 
In the limit in which the source 
is small compared to the value of the condensate, both the $U(1)$ symmetry and rotations are broken pseudo-spontaneously. This limit has been extensively studied using holography together with hydrodynamics and effective field theory in the context of translations (see \cite{RevModPhys.95.011001} for a review). It is commonly discussed in the case of chiral symmetry, \underline{i.e.} pions \cite{Grossi:2021gqi,Cao:2022csq}, and it has been recently considered for the simpler case of a single U(1) global symmetry \cite{Ammon:2021slb,Delacretaz:2021qqu,Armas:2021vku}. In the opposite regime, in which the source is parametrically larger than the vector condensate, the physics 
should be controlled by the mechanism of explicit symmetry breaking. 

One of the main results of our analysis is that the explicit breaking of rotations leads to a \emph{suppression} of $\eta/s$ at small temperatures, as compared to its behavior in the purely spontaneous case. 
This confirms our intuition that the two mechanisms of symmetry breaking compete against each other at small temperatures.
In addition, by independently tuning the effects of explicit and spontaneous symmetry breaking, we prove that broken rotational invariance by itself
does \emph{not} necessarily imply the violation of the KSS bound. 
Indeed, we find that in this model, even in the limit in which the breaking is mostly explicit, the $\eta/s$ ratio does not go below the KSS value, $1/4\pi$. 
Moreover, we find that such a ratio always reaches a constant value, which nevertheless depends on the source of explicit symmetry breaking, in the limit of small temperature. This behavior, which is different from the cases discussed before in the literature, has to be ascribed to the properties of the near-horizon geometry in the extremal limit, which becomes a mild anisotropic deformation of AdS$_5$, which we label \textit{deformed AdS$_5$}. Ultimately, the fate of $\eta/s$ in anisotropic systems depends crucially on the RG flow properties of the operator responsible for the breaking of rotations, which could give rise to a complex landscape of scenarios. This is also relevant to the issue of a potential lower bound on $\eta/s$. Given that the mechanisms for  spontaneous and explicit rotational symmetry breaking  push $\eta/s$ in different directions, it is natural to wonder whether their competing effects could lead to a minimum for $\eta/s$. 

The lesson we draw from our analysis is that, while in principle these combined effects could be used to generate such a minimum, doing so would require a much more drastic deformation of the IR geometry at extremality. It would also entail a delicate balancing between the different mechanisms at play, and it is hard to see how this could be of a universal nature.
This of course would not be related to the minimum appearing in real fluids, which is due to the liquid to gas transition, and is not linked to any symmetry breaking pattern. On the contrary, a possible application of our results could be found in the context of nematic liquids or, more generally, nematic liquid crystals. There, momentum transport is strongly anisotropic and the viscosities, which are classified using the notations introduced by Miesowicz \cite{MIESOWICZ1946}, can be measured experimentally and show an interesting temperature dependence and small values \cite{JanJadzyn_2001,diogo:jpa-00209451,Chen:15}. The behavior of the viscosity at the nematic/isotropic transition has also been experimentally investigated \cite{D1CC06111A}.
Finally, it would be valuable to explore the implications of our results for the physics of the strongly coupled quark gluon plasma, where the flow is anisotropic and the precise temperature dependence of $\eta/s$ is expected to play a key role in shedding light on the dynamics near the QCD phase transition.  \\

\section{The holographic Setup}
 
We work with a five-dimensional holographic model of a p-wave superfluid, describing gravity coupled to $SU(2)$  Yang-Mills vector fields in a spacetime asymptotic to AdS \cite{Gubser:2008wv}. 
We take the action to be 
\begin{equation}
\label{eq:action}
S = \int\!d^5x\,\sqrt{-g} \, \left [ \frac{1}{2\kappa_5^2} \left( R + \frac{12}{L^2}\right) - \frac{1}{4\hat g^2} \, F^a_{MN} F^{aMN} \right] + S_{\text{bdy}}\,, 
\end{equation}
where $\kappa_5$ is the five-dimensional gravitational constant,
$L$ the AdS radius and $\hat g$  the Yang-Mills coupling constant (we follow the notation of \cite{Erdmenger:2011tj}). The boundary action $S_{\text{bdy}}$ includes the  Gibbons-Hawking boundary term for a well-defined Dirichlet variational principle and a surface counterterm for removing divergence (see Appendix~\ref{app:renorm}).

The $SU(2)$ field strength $F^a_{MN}$ is
\begin{equation}
F^a_{MN}=\partial_M A^a_N -\partial_N A^a_M + \epsilon^{abc}A^b_M A^c_N \,,
\end{equation}
where $ A^a_M $ are the components of the matrix valued gauge field  $ A = A^a_M \tau^a dx^M $, with $\tau^a$ the $SU(2)$ generators, and $\epsilon^{abc}$ the three dimensional Levi-Civita tensor. 
The corresponding Einstein and Yang-Mills equations are then 
\begin{align}
\label{eq:einsteinEOM}
R_{M N}+\frac{4}{L^2}g_{M N}&=\kappa_5^2\left(T_{MN}-\frac{1}{3}{T_{P}}^{P}g_{MN}\right)\,, \\
\label{eq:YangMillsEOM}
\nabla_M F^{aMN}&=-\epsilon^{abc}A^b_M F^{cMN} \,,
\end{align}
where the Yang-Mills stress-energy tensor $T_{MN}$ is
\begin{equation}
\label{eq:energymomentumtensor}
T_{M N}=\frac{1}{\hat{g}^2}\left(F^a_{PM}{F^{aP}}_{N}-\frac{1}{4}g_{MN} F^a_{PQ}F^{aPQ}\right)\,.
\end{equation}
We begin with the ansatz
\begin{equation}\label{ansatz}
      ds^2=-u(r)dt^2+\frac{1}{u(r)}dr^2+h(r)dx^2+v(r)(dy^2+dz^2)\,.
\end{equation}
\begin{equation}
A=\phi(r)\tau^3 d t+w(r)\tau^1 d x\,,\nonumber
\end{equation}
where the AdS boundary is at $r\rightarrow\infty$ and the event horizon is located at $r=r_h$ with $u(r_h)=0$.

The background equations of motion read
\begin{subequations}\label{bkeoms}
\begin{align}
    0&=u'\left(\frac{h'}{2h}+\frac{v'}{v}\right)+u\left(\frac{h' v'}{h v}+\frac{v'^2}{2v^2}-\frac{\alpha^2 w'^2}{h}\right)-\frac{\alpha^2 \phi^2 w^2}{h u}+\alpha^2 \phi'^2-12\,,\\
    0&=h'' -\frac{h'^2}{2h}+h'\left(\frac{u'}{u}+\frac{v'}{v}\right)-h\left(\frac{8}{u}-\frac{2\alpha\phi'^2}{3u}\right)+\frac{4\alpha^2}{3}\left(w'^2-\frac{\phi^2 w^2}{u^2}\right)\,,\\
    0&=v'' +v'\left(\frac{h'}{2h}+\frac{u'}{u}\right)-v\left(\frac{8}{u}-\frac{2\alpha^2\phi'^2}{3u}-\frac{2\alpha^2\phi^2w^2}{3hu^2}+\frac{2\alpha^2w'^2}{3h}\right)\,,\\
    0&=\phi'' + \phi'\left(\frac{h'}{2h}+\frac{v'}{v}\right)-\frac{\phi w^2}{hu}\,,\\
    0&=w''-w'\left(-\frac{h'}{2h}+\frac{v'}{v}+\frac{u'}{u}\right)+\frac{\phi^2 w}{u^2}\,,
\end{align}
\end{subequations}
where primes are derivatives with respect to $r$ and 
we have introduced a new parameter,
\begin{equation}
    \alpha \equiv \frac{\kappa_5}{\hat g}\,.
\end{equation}
We have also set $L=1$. Note that when $w(r)$ vanishes,  $h(r)=v(r)$ and the solutions have $SO(3)$ rotational invariance.
However, 
backgrounds with non-zero $w(r)$ preserve only $SO(2)$ symmetry  along the $y,z$ directions.

In what follows we will discuss the computation of the shear viscosities in this system, for the case in which rotational invariance is broken spontaneously as well as explicitly. 
The explicit symmetry breaking case will be realized by ensuring that the gauge field component $A_x^1 = w(r)$ has a constant mode (\emph{i.e.}, it is sourced).

For completeness, we include the 
form of the background near the horizon and the boundary, which will be needed to compute the shear viscosity.
Near the horizon, the background fields take the following form,
\begin{equation}\label{IR}
\begin{split}
u &= 4 \pi T (r -r_h)+ ... \,,\\
v &= v_1 +\frac{v_1 (12-\alpha^2\phi_2^2)}{6\pi T}(r -r_h)+... \,,\\
h &= h_1 +\frac{h_1 (12-\alpha^2\phi_2^2)}{6\pi T}(r -r_h)+...\,,\\
w &= w_1 + \mathcal{O}\left((r-r_h)^2\right)\,,\\
\phi &=\phi_1(r -r_h)+... \,,
\end{split}
\end{equation}
where $r_h$ denotes the black hole horizon and $v_1$, $h_1$, $w_1$ and $\phi_1$ are free coefficients. Note that we have imposed the regularity condition that $A_t=\phi$ should vanish at the horizon. The boundary expansion is cumbersome and its full expression is given in Appendix \ref{app:renorm}. Schematically, it reads
\begin{equation}
\begin{split}
    u(r)=&r ^2+\ldots+
    \frac{u_{{b1}}}{r ^2}\ldots,\\
    v(r)=&r^2+\ldots+\frac{v_{{b1}}}{r ^2}+\ldots,\\
    h(r)=&r^2+\ldots - \frac{  \frac{1}{6} \alpha ^2 \mu ^2 w_{{b0}}^2 + 2 v_{{b1}}}{r ^2}+ \ldots,\\
    w(r)=& w_{{b0}}+\ldots+\frac{ w_{{b1}}}{r ^2}+ \ldots,\\
    \phi(r)=&\mu+\ldots+\frac{\phi _{{b1}}  }{r ^2}+ \ldots,
 \end{split}
\end{equation}
where the coefficients which are not displayed are determined by $\{u_{b1}, v_{b1}, w_{b0}, w_{b1}, \mu, \phi_{b1}\}$. Here, $\mu$ is the chemical potential and $w_{b0}$ is the source that explicitly breaks the rotational invariance. When $w_{b0}=0$, the rotational symmetry can still be broken spontaneously below a certain critical temperature $T=T_c$. 

Using holographic renormalization, we then obtain the expectation value of the energy-momentum tensor, the current density and the charge density of the boundary theory,
\begin{equation}\label{thermodynamics}
\begin{split}
    \mathcal{E}=\langle T_{tt\rangle}&=-\frac{9 u_{b1}+2 \alpha^2 w_{b0}^2\mu^2}{6\kappa_5^2},\\
    \mathcal{P}_{\|}=\langle T_{xx}\rangle&=-\frac{u_{b1}+8v_{b1}}{2\kappa_5^2},\\
    \mathcal{P}_{\bot}=\langle T_{yy}\rangle&=T_{zz}=\frac{\alpha^2  \mu ^2 w_{b0}^2-6 u_{b1}+24 v_{b1}}{12 \kappa _5^2},\\
    \langle J_1^x \rangle&=\frac{\alpha^2  \left(4 w_{b1}-\mu ^2 w_{b0}\right)}{2 \kappa _5^2},\\
   \mathcal{\rho}= \langle J^t_3\rangle&=-\frac{\alpha^2  \left(\mu  w_{{b0}}^2+4 \phi _{{b1}}\right)}{2 \kappa _5^2},
\end{split}
\end{equation}
while other components vanish.  More details on the holographic renormalization procedure can be found in Appendix~\ref{app:renorm}. It is clear that the source $w_{b0}$ has a non-trivial contribution to the above thermodynamic quantities. In the presence of the source, the pressure longitudinal to the condensate $\mathcal{P}_{\|}$ is different from the one perpendicular to the condensate $\mathcal{P}_{\bot}$. This is tantamount to saying that isotropy is broken in an explicit way, at the level of the UV action.

Thanks to the scaling symmetry of the system, one can obtain a radially conserved charge~\cite{Cai:2021obq}
\begin{equation}
\mathcal{Q}(r)=\frac{1}{2\kappa_5^2} v^2\sqrt{h} \left[\left(\frac{u}{v}\right)'-\frac{2\alpha^2}{v}\phi\phi'\right]\,.
\end{equation}
One can also check that $\mathcal{Q}'(r)=0$ by directly substituting the equations of motion~\eqref{bkeoms}. Evaluating $\mathcal{Q}$ at the horizon $r=r_h$, where $u(r_h)=0$, we find 
\begin{equation}
\mathcal{Q}=T s\,,
\end{equation}
with $s=2\pi \sqrt{h(r_h)}v(r_h)/\kappa_5^2$ the entropy density of the black hole. If we evaluate $\mathcal{Q}$ at the AdS boundary, we obtain
\begin{equation}
\mathcal{Q}=\mathcal{E}+\mathcal{P}_{\bot}-\mu\, \rho\,.
\end{equation}
Then, using that $\mathcal{Q}'=0$, we obtain the expected Smarr thermodynamic relation 
\begin{equation}
\mathcal{E}+\mathcal{P}_{\bot}=T s+\mu\, \rho\,.
\end{equation}
Furthermore, the trace of the energy-momentum tensor reads
\begin{equation}
\langle{T_\mu}^\mu \rangle=-\mathcal{E}+\mathcal{P}_{\|}+2\mathcal{P}_{\bot}=\frac{\alpha^2\mu^2}{2\kappa_5^2}w_{b0}^2\,,
\end{equation}
which is positive in the presence of source, implying that conformal symmetry is broken.

\section{Shear Viscosity}

The universal behavior of $\eta/s$ in isotropic holographic models follows from the shear mode transforming as
a helicity two state under the rotational symmetry and 
decoupling from the remaining fluctuations, 
behaving as a massless scalar. The remarkably simple behavior $\eta/s=1/4\pi$ can be traced to the universality of its coupling.
This is no longer the case when the rotational symmetry is broken and the fluid is anisotropic. 
The viscous properties of the fluid are now described by a tensor, and -- while the helicity two mode is still universal -- additional shear modes are present, which can be non-universal and temperature dependent.

To compute the viscosities, the metric and $SU(2)$ vector fields must be perturbed appropriately. 
In the symmetry broken case, the fluctuations, which generically take the form
\begin{equation}
\delta g_{\mu\nu} = h_{\mu\nu}(x^\mu,r) \, e^{-i\omega t} \, , \quad  
\delta A^a_\mu = a^a_\mu(x^\mu,r)\, e^{-i\omega t} \, ,
\end{equation}
 can be classified according to how they transform under the $SO(2)$ symmetry 
 (for a detailed 
 discussion see e.g. \cite{Erdmenger:2011tj}).
 Ignoring the helicity zero sector, which does not contribute to the shear viscosities, the remaining modes can be divided as follows,
 \begin{itemize}
    \item helicity two: $h_{yz}, h_{yy}-h_{zz}$\,,
    \item helicity one: $ h_{xy}, h_{ty},  a_y^a$ \; (a=1,2,3)\,.
\end{itemize}
%
It is the helicity two perturbation $h_{yz}$ which leads to the universal 
$\eta_{yz}/s=1/4\pi$ result expected for isotropic systems. 
On the other hand, 
the helicity one mode $h_{xy}$ 
is responsible for a 
 non-universal shear viscosity $\eta_{xy}$. 
In our analysis we will focus exclusively on the helicity one sector, and refer the reader to \cite{Erdmenger:2011tj}
for a discussion of the helicity two case. Also, we will only consider the $\eta_{xy}$ viscosity which for simplicity will be denoted as $\eta$ in the rest of the manuscript.

In the helicity one sector, gauge-invariant perturbations are described \cite{Erdmenger:2011tj} by the combination
$\Psi = g^{yy}(\omega h_{xy}+k_{\|}h_{ty})$ and  $a_y^a$,
where $k_{\|}$ is the momentum longitudinal to the condensate (in this setup, along the $x$ direction).
Letting $\Psi_t=g^{yy}h_{ty}$ and $\Psi_x=g^{yy}h_{xy}$, one can see that $\Psi_t$ and $a^3_y$ decouple from the remaining helicity one modes, and obey
\begin{subequations}
\begin{align}
    \Psi_t'+\frac{2\alpha^2a_y^3 \phi'}{v}=0 \, ,\\
    a_y^{3\prime\prime}+a_y^{3\prime}\left(\frac{h'}{2h}+\frac{u'}{u}\right)+a_y^{3}\left(\frac{\omega^2}{u^2}-\frac{2\alpha^2\phi'^2}{u}-\frac{w^2}{hu}\right)=0 \, .
\end{align}
\end{subequations}
Since they don't contribute to the shear viscosity, we ignore them from now on.
The remaining perturbations $\Psi_x$, $a^1_y$ and $a^2_y$ obey
\begin{subequations}
\begin{align}
    \Psi_x''+\Psi_x'\left(\frac{u'}{u}-\frac{h'}{2h}+\frac{2v'}{v}\right)+\frac{2\alpha^2a_y^{1\prime}w'}{v}+\frac{\omega^2\Psi_x}{u^2}-\frac{2\alpha^2 a_y^1 \phi^2 w}{u^2v}+\frac{2i\omega\alpha^2 a_y^2 \phi w}{u^2v}=0 \, , \\
    a_y^{1\prime\prime}+a_y^{1\prime}\left(\frac{h'}{2h}+\frac{u'}{u}\right)-\frac{v\Psi_x^\prime w'}{h}+a_y^{1}\left(\frac{\omega^2}{u^2}+\frac{\phi^2}{u^2}\right)-\frac{2i\omega a_y^2 \phi}{u^2}=0 \, , \\
    a_y^{2\prime\prime}+a_y^{2\prime}\left(\frac{h'}{2h}+\frac{u'}{u}\right)+a_y^2\left(\frac{\omega^2}{u^2}+\frac{\phi^2}{u^2}-\frac{w^2}{hu}\right)+\frac{2i \omega a_y^1 \phi}{u^2}-\frac{i\omega \Psi_x v \phi w}{h u^2}=0 \, .
\end{align}
\end{subequations}
What makes the computation of the non-universal shear viscosity highly non-trivial, and typically requires numerics, is that these modes are all coupled to each other. 

However, as we show next, working perturbatively in the angular frequency $\omega$ will 
simplify the analysis considerably, and will allow us to obtain an analytic expression for the non-universal $\eta/s$ which depends only on the horizon structure of the background. 
We stress that an expansion in powers of the frequency is justified in this context because hydrodynamics 
is, after all, the long wavelength, low frequency description of the system. 
Given a shear perturbation which is sourced by $h_{xy}^{(0)}$, a source for the dual operator $T^{xy}$, the corresponding viscous response in linear response theory is given by:
\begin{equation}\label{ok}
    \delta \langle T^{xy}\rangle=-\eta \,\partial_t h_{xy}^{(0)} = i \omega\, \eta \,h_{xy}^{(0)}\,,
\end{equation}
where the source has been Fourier transformed (see \cite{Natsuume:2014sfa} for a pedagogical review of this derivation). Notice that $\partial_t h_{xy}^{(0)}$ is a shear strain rate\footnote{Fluids do not respond to a static shear strain.}. 
 To extract 
$\eta/s$ we will make use of 
Kubo's formula,
\begin{equation}\label{eqkubo}
 \eta=-\lim_{\omega\to 0}\ \frac 1\omega\,  \mathrm{Im}\, \mathcal{G}^R_{xyxy}(\omega, k=0)   \, ,
\end{equation}
where $\mathcal{G}^R_{xyxy}(\omega, k)$ is the retarded Green's function for the operator $T^{xy}$.
Therefore, it will suffice to expand the perturbations 
to linear order in $\omega$ (higher frequency terms in the expansion will not contribute to $\eta$).

\subsection{An Analytical Horizon Formula}

Following the strategy used in 
\cite{Landsteiner:2016stv} (see also \cite{Baggioli:2018bfa}), we 
expand the metric and gauge field perturbations $\Psi_x$, $a^1_y$, and $a^2_y$ in powers of frequency $\omega$,
\begin{subequations}
\begin{align}
    \Psi_{x}&=u^{-i\omega/(4\pi T)}\, \left(\Psi_{x}^{(0)} +\omega \, \Psi_{x}^{(1)}+ ...\right),\\
    a^1_{y} &= u^{-i\omega/(4\pi T)}\, \left(a_{y1}^{(0)} +\omega \, a_{y1}^{(1)}+ ...\right),\\
    a^2_{y} &= u^{-i\omega/(4\pi T)} \, \left(a_{y2}^{(0)} +\omega \, a_{y2}^{(1)}+ ... \right),
\end{align}
\end{subequations}
where for our purposes it is sufficient  to stop at linear order in $\omega$. 
The temperature-dependent prefactor is needed to ensure that the perturbations obey incoming wave boundary conditions at the horizon.
To zeroth order in $\omega$, the perturbation equations of motion are:
\begin{subequations}
\begin{align}
    0&=\Psi_x^{(0)''}+\Psi_x^{(0)'}\left(\frac{u'}{u}-\frac{h'}{2h}+\frac{2v'}{v}\right)+\frac{2\alpha^2a_{y1}^{{(0)}'}w'}{v}-\frac{2\alpha^2 a_{y1}^{{(0)}} \phi^2 w}{u^2v}\,,\\
    0&=a_{y1}^{(0)''}+a_{y1}^{(0)'}\left(\frac{h'}{2h}+\frac{u'}{u}\right)-\frac{v\Psi_x^{(0)\prime} w'}{h}+\frac{a_{y1}^{(0)}\phi^2}{u^2}\,,\\
    0&=a_{y2}^{{(0)}''}+a_{y2}^{{(0)}'}\left(\frac{h'}{2h}+\frac{u'}{u}\right)+a_{y2}^{(0)}\left(\frac{\phi^2}{u^2}-\frac{w^2}{hu}\right)\,.
\end{align}
\end{subequations}
Since we are interested in the shear viscosity, we turn off the source for gauge field perturbations. Therefore, the simplest solution of the equations above takes
\begin{equation}
    \Psi_{x}^{(0)}=1 \, , \quad 
    a_{y1}^{(0)}=a_{y2}^{(0)}=0\, .
\end{equation}
Plugging this choice into the $\cal{O}(\omega)$ equations of motion leads to a significant simplification, and gives 
\begin{subequations}
\begin{align}
\label{shear1}
     0&=\Psi_x^{(1)''}+\Psi_x^{(1)'}\left(\frac{u'}{u}-\frac{h'}{2h}+\frac{2v'}{v}\right)+\frac{2\alpha^2a_{y1}^{{(1)}\prime}w'}{v}-\frac{2\alpha^2 a_{y1}^{{(1)}} \phi^2 w}{u^2v}\nonumber\\
       &+\frac{i}{4\pi T}\left(\frac{h'u'}{2hu}-\frac{2v'u'}{vu}-\frac{u''}{u}\right)\,,\\
       \label{gauge1}
    0&=a_{y1}^{(1)''}+a_{y1}^{(1)'}\left(\frac{h'}{2h}+\frac{u'}{u}\right)-\frac{v\Psi_x^{(1)\prime} w'}{h}+\frac{a_{y1}^{(1)}\phi^2}{u^2}+\frac{i}{4 \pi T}\frac{vu'w'}{hu}\,,\\
    0&=a_{y2}^{{(1)}''}+a_{y2}^{{(1)}'}\left(\frac{h'}{2h}+\frac{u'}{u}\right)+a_{y2}^{(1)}\left(\frac{\phi^2}{u^2}-\frac{w^2}{hu}\right)-\frac{i v \phi w}{hu^2}\,.
\end{align}
\end{subequations}
Note that the $a_{y2}^{(1)}$ perturbation has decoupled from the other two fluctuations, and can therefore be ignored. From now on, we 
will restrict our attention to the 
two coupled differential equations for 
$\Psi_x^{(1)}$ and $a_{y1}^{(1)}$.

After some manipulations, it is straightforward to show that (\ref{shear1}) can be solved by 
writing the shear perturbation $\Psi_x^{(1)}$ 
in the following integral form,
 \begin{equation*}
   \Psi_x^{(1)}(r)=\int_{r_h}^r \left[ \frac{i}{4\pi T}\frac{u'}{u}-\frac{2 \alpha^2 a_{y1}^{(1)} w'}{v} + C(\tilde{r}) \right] d \tilde{r} \, ,
    \end{equation*}
    where $C(r)$ is a function which must obey the following constraint, 
\begin{equation*}
         \left(\frac{2v'}{v}-\frac{h'}{2h}+\frac{u'}{u}\right)C(r)+C'(r)=0 \, .
    \end{equation*}
    The latter can be easily solved and yields, upon requiring that the shear perturbation 
    $\Psi_x^{(1)}$
    is regular at the horizon, the expression
     \begin{equation*}
    C(r)= -\frac{iv_1^2}{\sqrt{h_1}} \left(
    \frac{\sqrt{h}}{uv^2} \right) 
    \, ,
    \end{equation*}
where $v_1$ and $h_1$ are parameters that characterize the horizon expansion of the background, see~\eqref{IR}.
Finally, putting all these ingredients together we find 
\begin{equation}
\label{finalshear}
     \Psi_x^{(1)}(r)=\int_{r_h}^r \left[ \frac{i}{4\pi T}\frac{u'}{u}-\frac{2 \alpha^2 a_{y1}^{(1)} w'}{v}-\frac{i v_1^2}{\sqrt{h_1}} \frac{\sqrt{h}}{uv^2}\right] d \tilde{r}\,, 
    \end{equation}
an integral expression for the shear mode in terms of the background and the gauge field perturbation.

Now that we have ensured that the mode is well behaved near the horizon, we can  examine its boundary expansion. 
Recalling that $\Psi_{x}= g^{yy}h_{xy}$, we write the full perturbation to first order in the frequency, 
\begin{equation}
     h_{xy}=v(r)u(r)^{-i\omega/(4\pi T)}(\Psi_{x}^{(0)} +\omega \Psi_{x}^{(1)}+ ...)\,.
\end{equation}
Using the expressions for $\Psi_{x}^{(0)}$ and $\Psi_{x}^{(1)}$ obtained above, we have
\begin{equation}
\label{fullhxy}
    h_{xy}=v(r)u(r)^{-i\omega/(4\pi T)}\left(1+\omega\int_{r_h}^r \left[ \frac{i}{4\pi T}\frac{u'}{u}-\frac{2 \alpha^2 a_{y1}^{(1)} w'}{v}-\frac{i v_1^2}{\sqrt{h_1}} \frac{\sqrt{h}}{uv^2}\right] d\tilde{r}
    \right)\,.
\end{equation} 
The crucial next step is to obtain an approximate expansion for the integral that is valid near the boundary, from which to extract 
the retarded Green's function.
To do so, our strategy is going to be to Taylor expand the integral (\ref{fullhxy}) about the boundary, making use of the asymptotic expansions of the background components $\{u,v,h,w\}$ and of the gauge field perturbation $a_{y1}^{(1)}$. 
From the resulting boundary expansion of the shear fluctuation $h_{xy}$, it is then straightforward to read off the vev and source of its dual operator, 
and extract the retarded Green's function 
$\mathcal{G}^R_{xyxy} = \frac{\text{vev}}{\text{source}}$.
We refer the reader to Appendix \ref{two} for the details of the calculation, and here state the final results.
Using Kubo's formula~\eqref{eqkubo}, we can extract the shear viscosity, 
\begin{equation}\label{shearmain}
    \eta=
    \frac{1}{2\kappa_5^2}\frac{ v_1^2}{\sqrt{h_1}}\,.
\end{equation}
Combining this result with the expression for the entropy density, 
$s=\frac{2\pi}{\kappa_5^2}\sqrt{h_1}v_1$,
we finally obtain the ratio
\begin{equation}
\label{res}
\boxed{
    \frac{\eta}{s}=\frac{1}{4\pi}\frac{v_1}{h_1}}\,.
\end{equation}
Eq.\eqref{res} is independent of whether rotations are broken explicitly or spontaneously and it coincides with the well-know formula (see for example \cite{Jain:2015txa}) for anisotropic systems given by
\begin{equation}
    \frac{\eta}{s}=\frac{1}{4\pi}\,\frac{g_{zz}}{g_{xx}}\Big|_{r_h}\,,
\end{equation}
where $r_h$ is the location of the horizon.

\subsection{Numerical Analysis}
\begin{figure}
    \centering
    \includegraphics[width=0.32\textwidth]{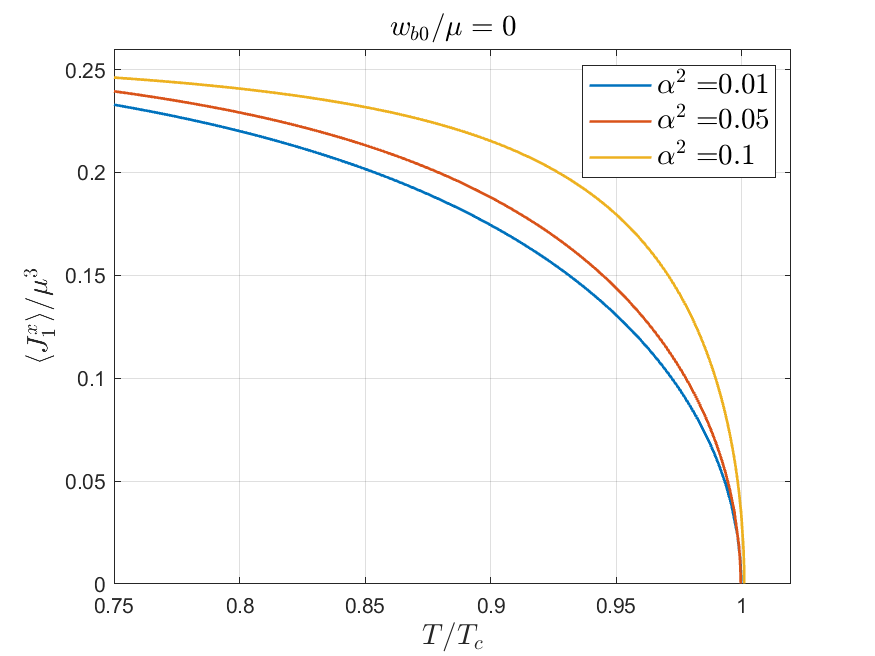}
    \includegraphics[width=0.32\textwidth]{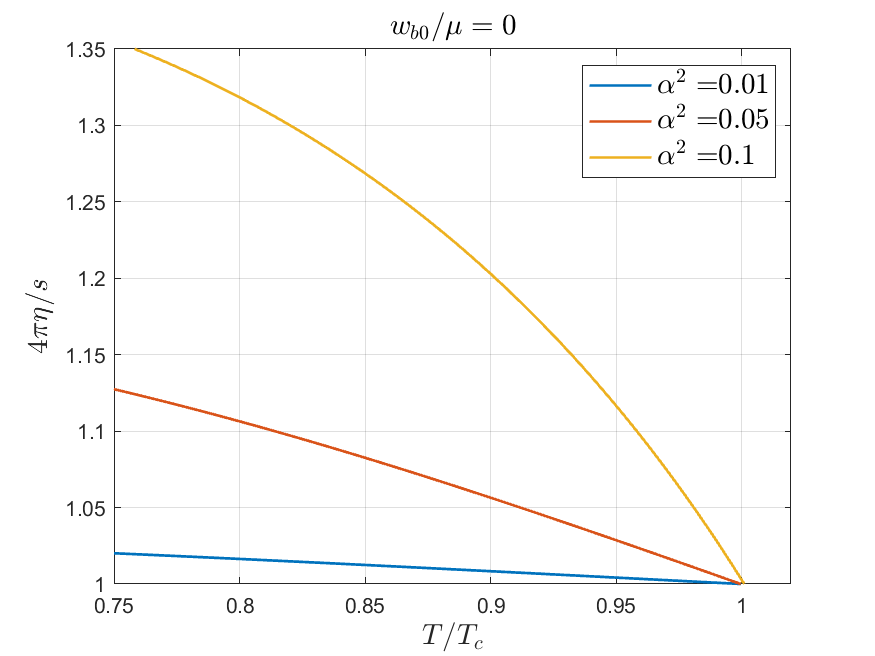}
    \includegraphics[width=0.32\textwidth]{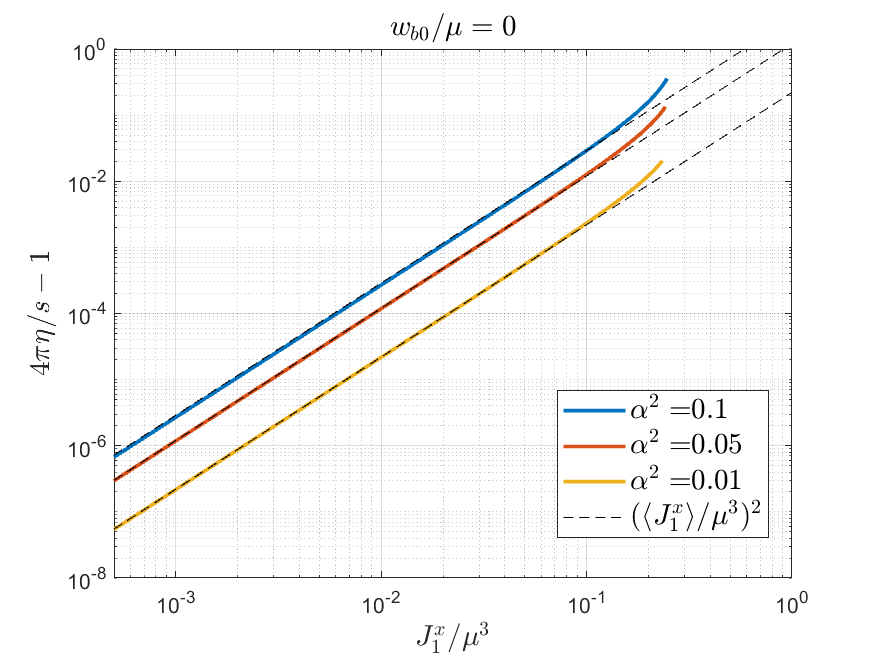}
    
    \caption{\textbf{Left:} Temperature dependence of the condensate $\langle J_1^x \rangle$ in the purely spontaneous case. \textbf{Center:} Temperature dependence of $\eta/s$ in the purely spontaneous case. \textbf{Right:} Double-logarithmic plot of the deviation $4\pi \eta/s-1$ as a function of the spontaneous condensate $\langle J_1^x\rangle$ close to the critical point $T=T_c$. The dashed lines guide the eyes towards the universal scaling $\sim \langle J^x_1\rangle ^2$.}
    \label{fig:todo}
\end{figure}

Having obtained the analytical 
horizon formula (\ref{res}),  we can test its validity numerically. 
Indeed, the numerics make use of the full numerical background, and thus provide a non-trivial check of our horizon formula. As we will see, we find excellent agreement between the two methods.
Before moving to our new results, however, we find it instructive to revisit the findings in the purely spontaneous case ($w_{b0}=0$) reported in \cite{Erdmenger:2011tj}. 
In the left panel of Fig.\ref{fig:todo}, we show the behavior of the vector condensate $\langle J_1^x \rangle$ as a function of the reduced temperature $T/T_c$ for different values of the coupling $\alpha$. This shows clearly that the system is undergoing a phase transition at $T=T_c$. If the coupling is smaller than a certain critical value $\alpha_c\approx 0.365$ \cite{Erdmenger:2011tj}, then the phase transition is of second order and the condensate follows the mean field scaling
\begin{equation}
    \langle J^x_1\rangle \propto \left(T_c-T\right)^{1/2}\,,
\end{equation}
as shown in the left panel of Fig.\ref{fig:todo}. The condensate also grows monotonically by increasing the coupling parameter $\alpha$.

In the normal phase, $T>T_c$, the viscosity saturates the KSS limit:
\begin{equation}
    \frac{\eta}{s}=\frac{1}{4\pi},\qquad T>T_c\,.
\end{equation}
In the broken phase, $T<T_c$, (see central panel of Fig.\ref{fig:todo}) the $\eta/s$ ratio grows with decreasing temperature and acquires a non-universal value that strongly depends on the value of $\alpha$. While our results for the non-universal $\eta/s$ are in qualitative agreement with those of \cite{Erdmenger:2011tj}, we find a difference in the temperature dependence, which becomes more apparent as the temperature is lowered well below $T_c$. We believe that the discrepancy may be explained by the different numerical precision -- in our analysis we took values of $\omega$ that are very close to zero. Working with larger values of $\omega$ seems to yield results that are closer to those of \cite{Erdmenger:2011tj}. In other words, this implies that the results of \cite{Erdmenger:2011tj} are not completely capturing the $\omega \rightarrow 0$ limit needed to define the shear viscosity coefficient via the corresponding Kubo formula.
\begin{figure}
    \centering
    \includegraphics[width=0.48\textwidth]{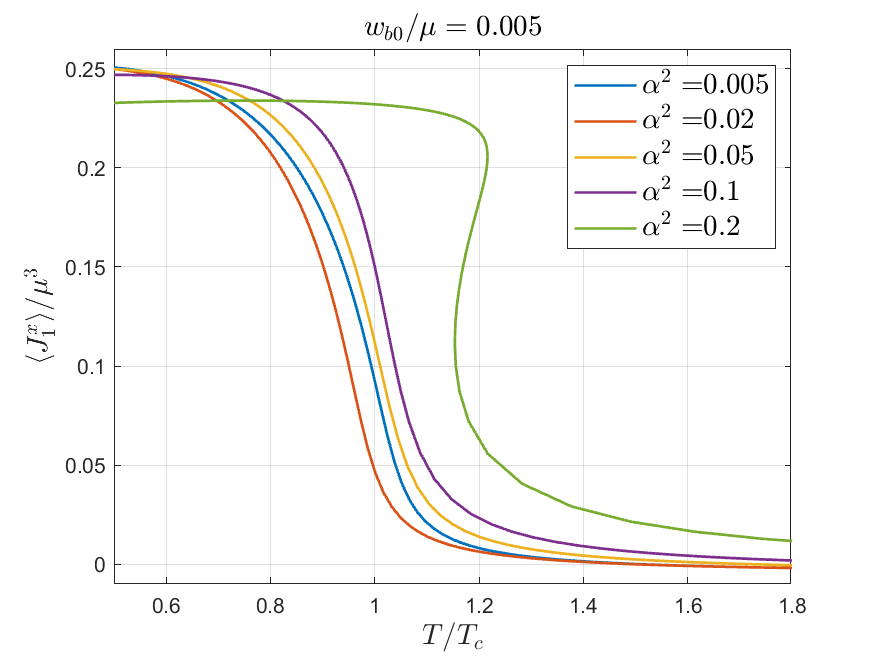}
    \includegraphics[width=0.48\textwidth]{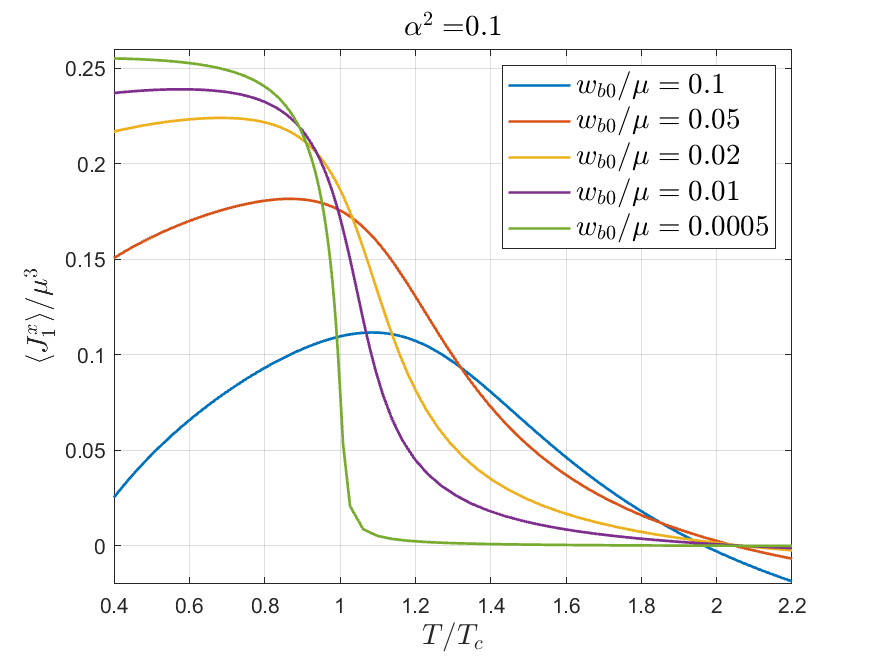}
    \caption{\textbf{Left: }The current expectation value $\langle J^x_1\rangle$ as a function of the reduced temperature $T/T_c$ for a fixed value of the source $w_{b0}/\mu=0.005$, changing the strength of the coupling $\alpha^2$. The green curve, $\alpha^2=0.2$, corresponds to a first-order phase transition. \textbf{Right: }The current expectation value $\langle J^x_1\rangle$ as a function of the reduced temperature $T/T_c$ for a fixed value of the coupling $\alpha^2=0.1$, changing the strength of explicit breaking source $w_{b0}/\mu$.}
    \label{fig:2}
\end{figure}

Interestingly, we notice that close to the critical point, the deviation of the $\eta/s$ ratio from the ``universal'' KSS value is well parameterized by the following phenomenological expression:
\begin{equation}\label{relrel}
    \left(\frac{4\pi\, \eta}{s}-1\right) \propto \langle J^x_1 \rangle^2\,.
\end{equation}
This result is not surprising and could probably be derived using a Ginzburg-Landau formalism, as done in the case of holographic supersolids in \cite{Baggioli:2022aft}.

We are now ready to consider the case in which a small source of explicit breaking of rotational invariance is added, $w_{b0}\neq 0$. In this limit of small source (compared to the value of the condensate), the breaking of rotational invariance is labelled as pseudo-spontaneous. The behavior of the condensate as a function of the reduced temperature is shown in Fig.\ref{fig:2}. 
In the left panel we vary the coupling, while the source is kept fixed at a small value.
In the right panel, instead, the coupling is held fixed while the source $w_{b0}$ is varied.

For small values of the source, the sharp critical behavior visible in Fig.\ref{fig:todo} is replaced by a smooth crossover, and the value of $\langle J_1^x \rangle$ is non-zero at any temperature. This behavior can be rationalized using Ginzburg-Landau theory and it has been observed already in several holographic models, including the cases of $U(1)$ symmetry \cite{Ammon:2021slb} and chiral symmetry \cite{Cao:2022csq}. Moreover, we see that for temperatures (roughly) below $T_c$, the condensate decreases when the source increases, while above $T_c$ the trend is exactly the opposite. This is the same qualitative behavior we observe in $\eta/s$, as we discuss next. Finally, we find that the turning point, defined as $\partial \langle J^x_1 \rangle/\partial T=0$,  moves towards larger temperature by increasing the value of the source $w_{b0}$.

\begin{figure}
    \centering
    \includegraphics[width=0.48\textwidth]{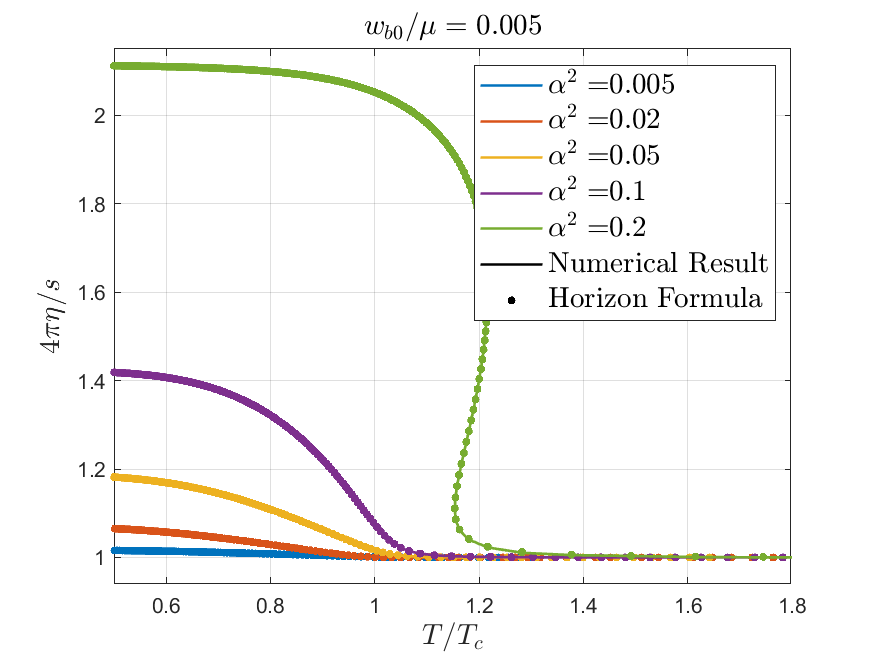}
    \includegraphics[width=0.48\textwidth]{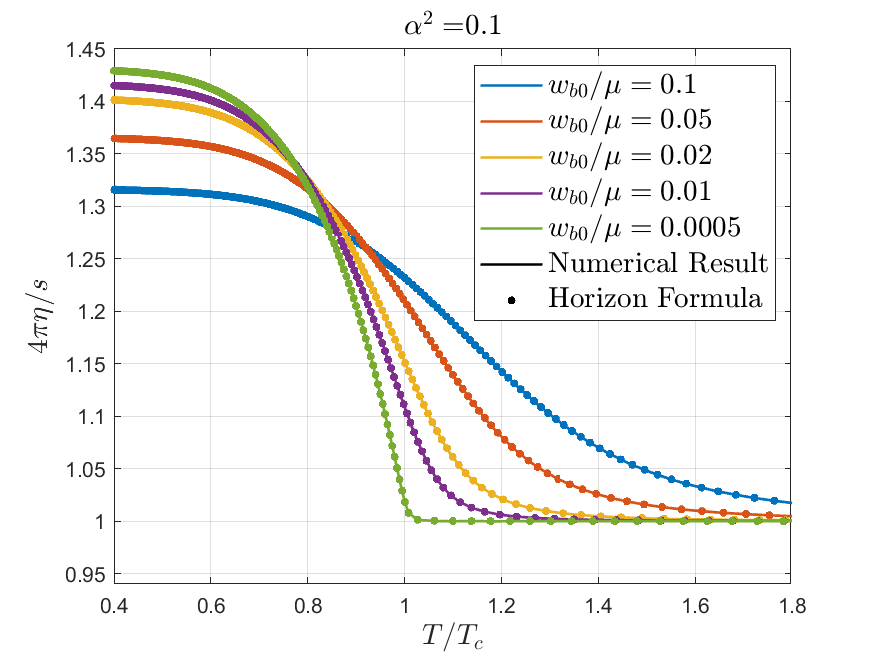}
    \caption{\textbf{Left: }The normalized $\eta/s$ ratio as a function of the reduced temperature for fixed source strength $w_{b0}/\mu$ and changing $\alpha^2$. The symbols are the numerical data while solid lines are our horizon formula, Eq.\eqref{res}. \textbf{Right: }A similar plot where we keep the coupling $\alpha^2$ fixed and change the strength of explicit symmetry breaking, $w_{b0}/\mu$.}
    \label{fig:3}
\end{figure}

Fig.\ref{fig:3} displays the temperature behavior of $\eta/s$ in the presence of a non-zero source $w_{b0}$ of explicit symmetry breaking. The numerical data (displayed with colored symbols) are in perfect agreement with the horizon formula, Eq.\eqref{res}, shown with solid lines.
In the left panel, the coupling $\alpha$ is varied and the explicit breaking scale $w_{b0}$ is held fixed, while in the right panel the situation is reversed.
We see clearly that the effect of a stronger coupling -- when  the source is small -- is to enhance the growth of $\eta/s$ towards small $T$.
The most significant result, on the other hand, is the suppression of $\eta/s$ towards smaller temperatures, as the source is increased.

Two features are notable.
First, that the temperature behavior of $\eta/s$ mimics that of the condensate.
Second, that there is a competition between spontaneous and explicit symmetry breaking.
Indeed, in the absence of a source of explicit symmetry breaking, $\eta/s$ grows towards small $T$, while when $w_{b0}$ is turned on, its effect is to suppress this growth.
Thus, the two different mechanisms of symmetry breaking are competing against each other. Interestingly, in the right panel of Fig.\ref{fig:3}, we observe a re-distribution in the profile of $\eta/s$ where the ``weight'' is transferred from temperatures below the critical one to temperatures above that. 
\begin{figure}
    \centering
    \includegraphics[width=0.48\textwidth]{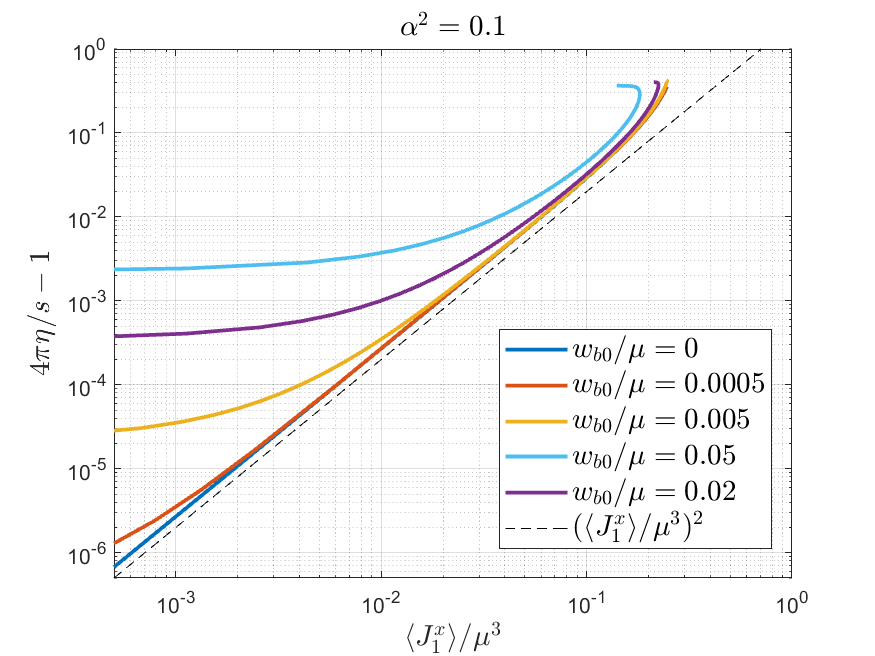}
     \includegraphics[width=0.48\textwidth]{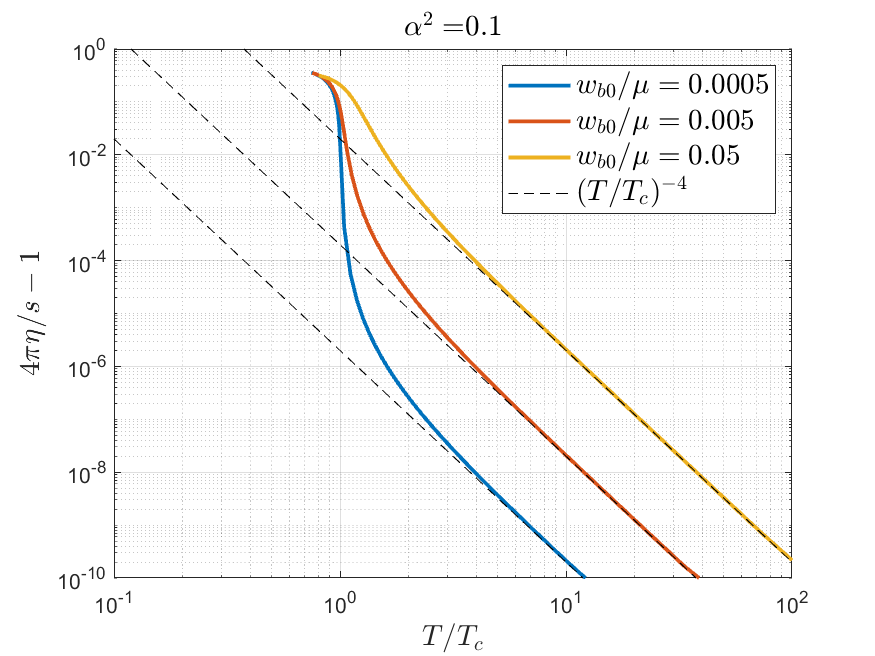}
    \caption{\textbf{Left: }A double-logarithmic plot showing the deviation from the spontaneous symmetry breaking relation $\left(\frac{4\pi\, \eta}{s}-1\right) \propto \langle J^x_1 \rangle^2$, Eq.\eqref{relrel}. Whenever this scaling, emphasized using a dashed black line, holds approximately, we can consider the system to be in the pseudo-spontaneous breaking regime. \textbf{Right: } A double-logarithmic plot displaying the power law decay of the deviation from the KSS bound, $4\pi\eta/s-1\sim T^{-4}$, for large temperatures and different values of the source. $T_{c0}$ indicates the critical temperature at zero source. The dashed lines guide the eyes towards the aforementioned scaling.}
    \label{fig:scaling_w0}
\end{figure}

To continue with our numerical analysis, in Fig.\ref{fig:scaling_w0} we examine in more detail the behavior of $\eta/s$ in presence of a small explicit symmetry breaking term. In the left panel, we show the deviation of the function $4\pi \eta/s-1$ from the scaling $\sim \langle J_1^x \rangle^2$ found in the purely spontaneous case, Eq.\eqref{relrel} (which is denoted by the dashed line in the plot). We find that, for small enough values of the source $w_{b0}$, the scaling still holds 
approximately, in the region around the critical point. On the contrary, for larger values of the explicit symmetry breaking parameter the scaling is completely lost. 
Thus, we can use this scaling region to establish whether the system can still be considered to be in the pseudo-spontaneous breaking regime or not. When the scaling regime is lost, no information of the spontaneous breaking remains, and the breaking of rotations becomes purely explicit. Additionally, in presence of a source $w_{b0}$, not only the condensate is non-zero at any finite value of temperature but also the difference $4 \pi \eta/s-1$, which parameterizes the deviation from the KSS bound. In the right panel of Fig.\ref{fig:scaling_w0}, we find that this difference vanishes as a power law $\sim T^{-4}$ at large temperature. It would be interesting to better understand 
the significance of this scaling behavior.

\section{On the (non) violation of the KSS bound}

\begin{figure}
    \centering
    \includegraphics[width=0.48\textwidth]{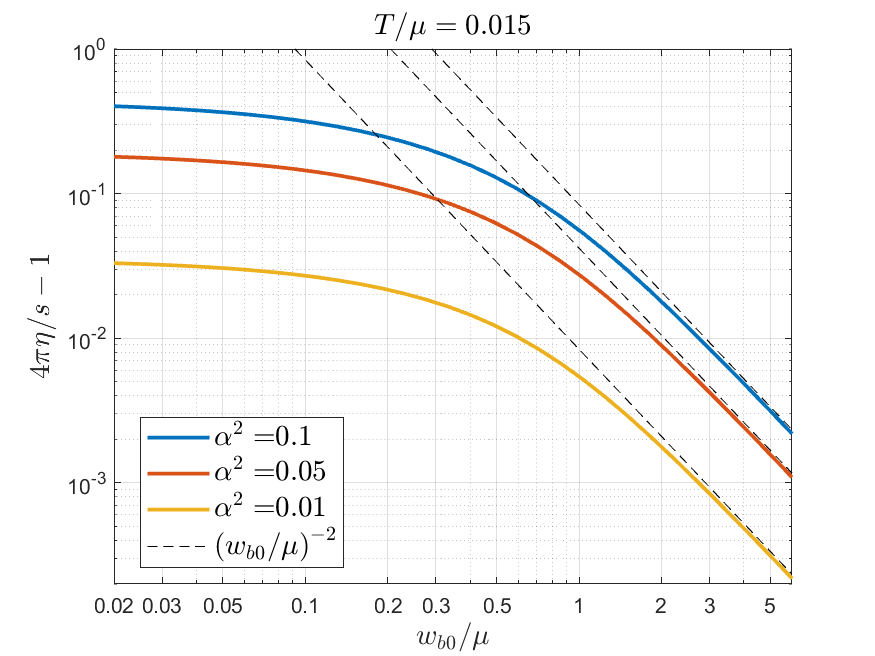}
    \includegraphics[width=0.48\textwidth]{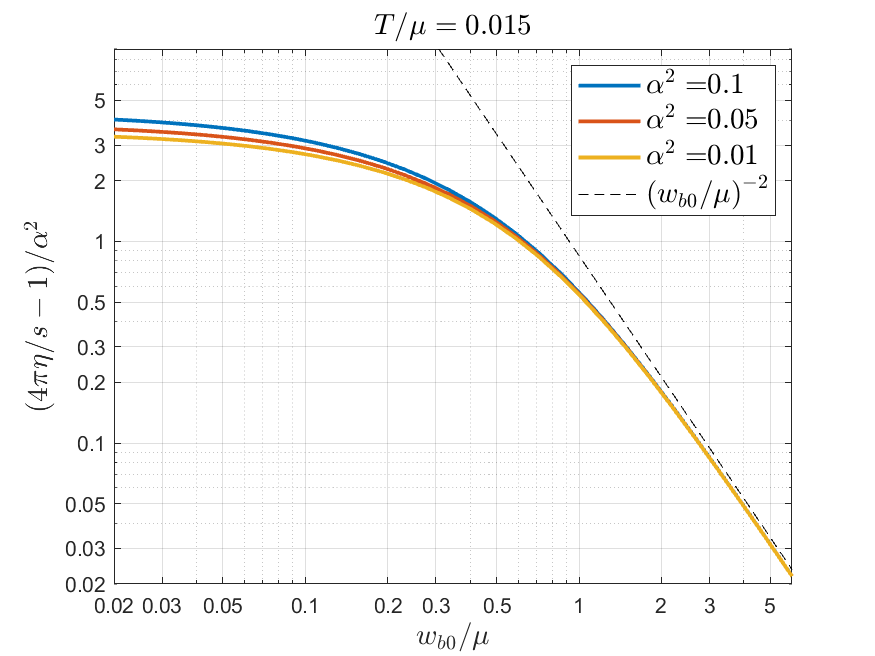}
    \caption{\textbf{Left: }A double-logarithmic plot of the deviation from the KSS bound $\left(4\pi\eta/s-1\right)$ at small $T$, as a function of the explicit source $w_{b0}$. \textbf{Right: }The same figure normalized by the coupling $\alpha$. In both panels, the dashed lines indicate the scaling $\sim w_{b0}^{-2}$.}
    \label{fig:proof}
\end{figure}

As shown in Fig.\ref{fig:3}, and already mentioned in the previous section, the introduction of a source
that breaks rotations explicitly  competes with the effects of purely spontaneous symmetry breaking and induces a suppression of $\eta/s$ at low temperature. 
Given these results, it is natural to ask whether 
a large amount of explicit symmetry breaking could lead to a violation of the KSS bound, at sufficiently low temperatures.
In a number of previous studies in the literature (\emph{e.g.}, \cite{Jain:2015txa,Finazzo:2016mhm}), it was shown that the explicit breaking of rotations (driven by a uni-directional axion field or a strong magnetic field) causes $\eta/s$ to vanish 
as $T\rightarrow 0$ following a power-law behavior, thus 
violating the KSS bound strongly. 

In our model we can tune the amount of explicit symmetry breaking and, by making the source $w_{b0}$ very large, we can reach the regime in which it dominates over the spontaneous one (see the left panel of Fig.\ref{fig:scaling_w0} for a criterion to estimate this transition). In such a limit, when the source $w_{b0}$ is much larger than the spontaneous condensate $\langle J_x^1\rangle$, the rotational symmetry is broken explicitly. Nevertheless, at least for the values explored in the right panel of Fig.\ref{fig:3}, a violation of the KSS bound is still not seen.

In order to clarify this point, in Fig.\ref{fig:proof} we plot the deviation from the KSS bound $\left(4\pi\eta/s-1\right)$ at a small temperature $T \approx 0$, as a function of the explicit symmetry breaking scale $w_{b0}$. We observe that the deviation becomes closer and closer to zero
for larger values of the source $w_{b0}$, indicating that even in the limit in which the explicit breaking of rotational symmetry is strong, $\eta/s$ will not violate the KSS bound. Moreover, we observe a power-law decay of the deviation 
$\left(4\pi\eta/s-1\right)$,
which scales as $\sim w_{b0}^{-2}$. Thus, we see that for very large values of the source, $\eta/s$ approaches the universal value $1/4\pi$ from above, without any indication of dipping below it. 
Importantly, as shown in the right panel of Fig.\ref{fig:proof}, such a behavior is independent of the value of the coupling $\alpha$, and therefore universal within our holographic model.

\begin{figure}
    \centering
    \includegraphics[width=0.48\textwidth]{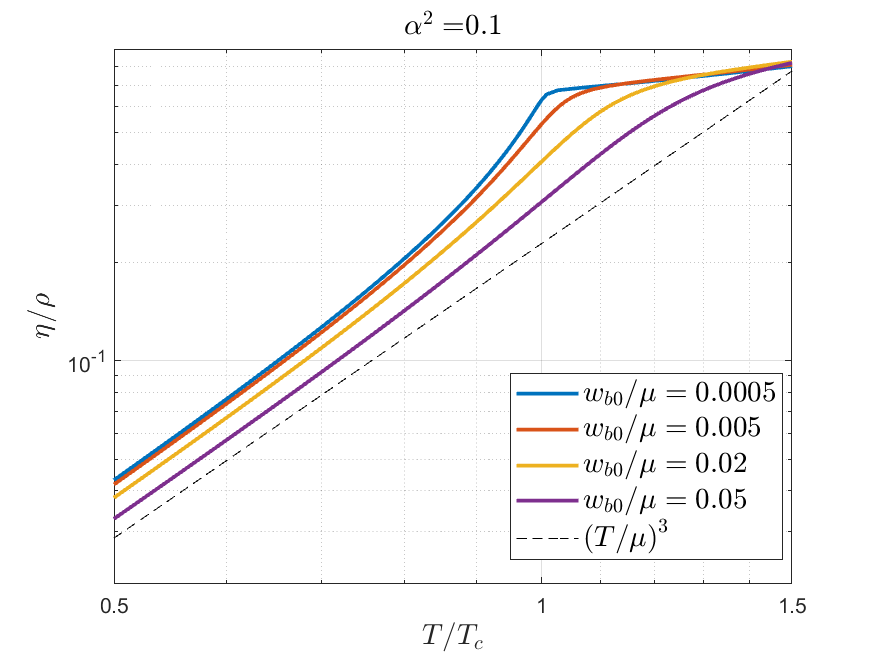}
    \includegraphics[width=0.48\textwidth]{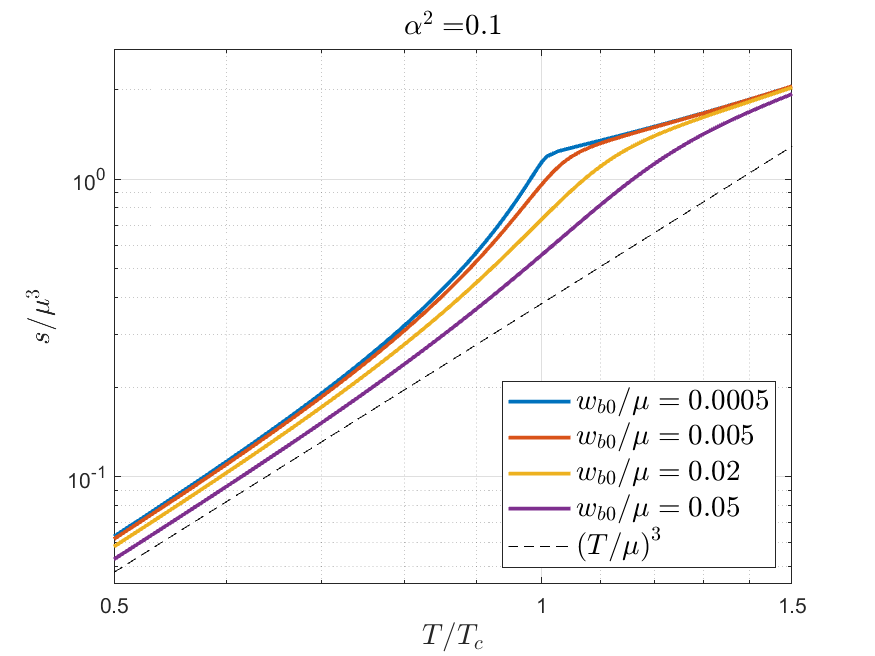}
    \caption{\textbf{Left: }Double-logarithmic plot of the $\eta/\rho$ ratio at small $T$ for different values of the source. \textbf{Right: }Double-logarithmic plot of the normalized entropy $s/\mu^3$ at small $T$. The dashed lines indicate the scaling behavior $\sim T^3$.}
    \label{fig:last}
\end{figure}

In order to understand the behavior of $\eta/s$ at low temperature better, we need to analyze in more detail the extremal near-horizon geometry, and the properties of the various geometrical, thermodynamical and transport properties therein. We start by plotting the normalized viscosity and entropy density as a function of temperature in Fig.\ref{fig:last}. As evident from the numerical data, both quantities scale as $\sim T^3$ in the deep IR. 
These scalings suggest that at zero temperature the IR geometry might be described by $AdS_5$, even in the presence of a source of explicit rotational symmetry breaking.
In other words, one would expect the gravitational solutions to be RG flows between an AdS$_5$ geometry in the UV and another AdS$_5$ geometry in the IR, very similar to the neutral Q-lattice models with broken translations in \cite{Hartnoll:2016tri}. As we will explicitly see, this is not exactly the case.

\begin{figure}
    \centering
    \includegraphics[width=0.55\textwidth]{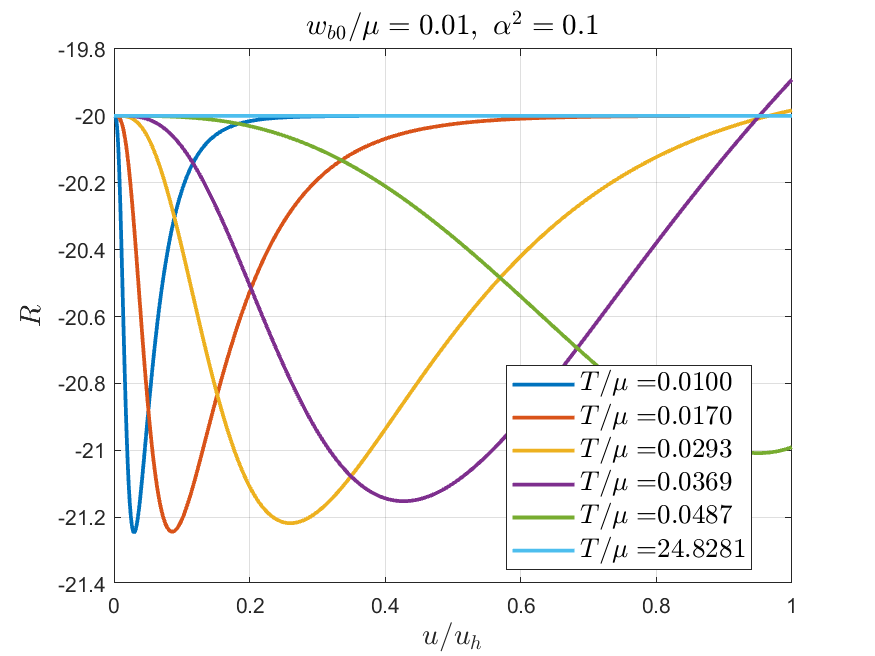}
    \caption{Ricci Scalar as a function of the normalized radial coordinate $u/u_h$ for different values of the normalized temperature. The value $R=-20$ corresponds to the AdS$_5$ geometry. We have introduced the new coordinate $u=1/r^2$ with $u_h=1/r_h^2$.}
    \label{RG}
\end{figure}

In order to confirm this, in Fig.\ref{RG} we plot the value of the Ricci scalar $R$ as a function of the normalized radial coordinate $u/u_h$ for different values of temperature. Here we have introduced the new coordinate $u=1/r^2$ with $u_h=1/r_h^2$.
The Ricci scalar in the UV, $u \rightarrow 0$, is given by the AdS$_5$ value $R=-20$. At low temperature we clearly observe that the same value is reached in the deep infrared, $u \rightarrow u_h$. 
This hints again at the fact that the 
near-horizon geometry in the near-extremal limit is AdS$_5$, as already suggested by the temperature scalings of the entropy density and the viscosity. 
Interestingly, the lower the temperature, the more the AdS$_5$ near-horizon geometry extends into the UV region. We might be tempted to conclude that the gravitational solutions are indeed ``boomerang'' RG flows between two AdS$_5$ geometries driven by an operator which breaks rotational invariance. This is not correct. Indeed, the latter operator still leaves an imprint on the IR AdS$_5$, leading to 
a metric of the schematic form, 
\begin{equation}\label{fin}
    ds^2 = - \alpha_t \, r^2 dt^2 + \frac{dr^2}{r^2} + r^2 \left(\alpha_x dx^2 + \alpha_z( dy^2 +  dz^2)\right) \, ,
\end{equation}
where the $\alpha_i$ are constants that depend on the particular 
value of $w_{b0}$. 
By looking at the spatial components in \eqref{fin}, one realizes that the metric is not exactly AdS$_5$, but it becomes such only after an anisotropic redefinition of the spatial coordinates. We will label the geometry in \eqref{fin} \textit{deformed AdS$_5$}. Notice that, even though the geometry shares many similarities with a standard AdS$_5$ spacetime (\emph{e.g.}, the value of the Ricci scalar), its isometries are profoundly different. In particular, the SO(3) symmetry in the $x,y,z$ coordinates is clearly broken to the SO(2) symmetry in the $x,y$ plane whenever $\alpha_z/\alpha_x \neq 1$.
In turn, this translates into different values of $\eta/s$ at extremality, corresponding 
to different choices of $w_{b0}$ (in terms of the coefficients above, we have $\eta/s \sim \alpha_z/\alpha_x$). In other words, rotational symmetry is not completely restored in the limit of small temperature,
where the imprints of the source are not vanishing.
This can be confirmed explicitly by looking at the value of the stress tensor components in the extremal limit, in presence of a source $w_{b0}$. In particular, one finds that $\langle T^{xx}\rangle\neq (\langle T^{yy}\rangle=\langle T^{zz}\rangle$). It is now clear why the behavior of $\eta/s$ at small temperatures in our setup is entirely different from the case with unidirectional axion fields or external magnetic fields. In those examples, the IR geometry remains strongly modified near extremality, becoming an AdS$_4$ $\times R$ geometry in the case of axion models \cite{Jain:2014vka} or a BTZ black hole times a two-dimensional torus in the case of a magnetic field \cite{Finazzo:2016mhm}, thus explaining
the different scaling of the $\eta/s$ ratio as $T\rightarrow 0$. 

Importantly, our analysis only implies that the $\eta/s$ ratio reaches a constant in the small temperature limit. Indeed, it does not reveal any information about the value that this constant
must take which, as shown above, depends strongly on the UV deformation of the theory. 
Interestingly, we see that the $\eta/s$ ratio at small temperature approaches the KSS bound value 
$1/4\pi$, in the limit of very strong source $w_{b0}\gg \mu$. This is reminiscent of the results in the neutral Q-lattice model of \cite{Hartnoll:2016tri} (see Fig.2 therein, where the strength of the source is indicated as $k/V$). There is nevertheless a big difference with our case. In the Q-lattice case, $\eta/s$ reaches the KSS value from below, while in ours from above. 
A possible way to understand this difference in more detail is rooted in the analysis of the slope of $\eta/s$ close to the critical point, at larger temperature. In our setup we always find a positive slope, while in the other cases the slope is negative. It would be interesting to understand which physical property of the operator responsible for the symmetry breaking determines the slope, and how one can derive it. A perturbative analysis close to the critical point might be helpful.

\section{Discussion}
In this work, we have revisited the computation of the shear viscosity to entropy ratio, $\eta/s$, in anisotropic holographic models. Our 
initial motivation was to 
better understand the 
different imprints left on $\eta/s$ by the spontaneous and explicit breaking of rotational invariance,   reported previously in the literature. In particular, it has been widely observed that an explicit anisotropy, induced for example by axion fields or an external magnetic field, would lead to a ``brutal" violation of the $\eta/s$ bound in the direction parallel to the anisotropy. Such a violation would persist up to zero temperature, close to where the $\eta/s$ ratio would decay following a power law behavior, $\left(T/\zeta\right)^2$, where $\zeta$ is the scale determining the anisotropy (\emph{e.g.}, the magnetic field $B$). 
On the contrary, in holographic systems with spontaneously generated  anisotropic phases (\emph{e.g.}, p-wave holographic superfluids), the universality of the $\eta/s$ ratio would  break down in a rather different way. Indeed, in these constructions it was observed that the $\eta/s$ ratio becomes larger than $1/4\pi$ in the broken symmetry phase and grows with the condensate. 
Thus, in these models the particular mechanisms for purely explicit vs. purely spontaneous symmetry breaking lead to 
sharply different behaviors. Note, however, that the translation symmetry is also broken in the former case.

In this paper, we have tuned the amount of explicit and spontaneous symmetry breaking, 
so that we could interpolate between the two. 
By doing so, we have seen that the presence of a source of explicit symmetry breaking leads to a suppression of $\eta/s$ at low temperature, as compared to the value it would have in the purely spontaneous case.
Interestingly, however, we have found that in the limit of large source we do not recover the behavior seen in the  holographic models with axions or external magnetic fields.
On the contrary, when the explicit symmetry  breaking is the dominant mechanism for the anisotropy, we observe that the $\eta/s$ ratio converges to a constant at small temperature, which is larger than the KSS value $1/4\pi$. We can explain this difference by looking at the nature of the near-horizon extremal geometry, which is intimately connected to the properties of the operator responsible for the breaking of rotations in the deep IR.

In conclusion, we find that knowledge of the rotational symmetry breaking pattern is not enough to understand the temperature dependence of $\eta/s$ in the symmetry broken phase. More directly, breaking rotational invariance explicitly does not necessarily imply a violation of the KSS bound, unlike what was observed  in previous studies in the literature.
The $\eta/s$ ratio is sensitive to further details of the symmetry breaking mechanism, in particular to the nature of the operator responsible for it, as in the case of holographic models with broken translations. Moreover, we found that the competition of spontaneous and explicit breaking of rotations is typically not enough to produce a minimum of the $\eta/s$ ratio as a function of the temperature, akin of that ubiquitously observed at the liquid-gas critical point. 
While it may be possible to use these competing effects to engineer a minimum, doing so would require properly balancing different effects, and perhaps fine tuning.

It would be interesting to perform a more detailed analysis of the quasinormal modes, as done for simpler holographic s-wave superfluids with \cite{Ammon:2021pyz} and without explicit symmetry breaking \cite{Amado:2009ts,Arean:2021tks}, or for anisotropic phases in \cite{Jain:2014vka}, and ascertain for example whether the momentum diffusion constant follows the $\eta/s$ behavior or not. For systems with broken translations, the answer is no \cite{Baggioli:2020ljz}. It would be also fruitful to consider other holographic models with broken rotations, as for example those presented in \cite{Cremonini:2014pca,Cai:2013aca,Garbayo:2022pqp,Hoyos:2020zeg}, or systems under shear \cite{Ji:2022ovs,Baggioli:2020qdg}, in order to reach a complete picture of the whole landscape of anisotropic phases. 

Finally, one fundamental question is left to be understood, which is what determines the value of the constant $\eta/s$ ratio at zero temperature in our holographic model, and how that depends on the properties of the dual field theory. In this direction, a perturbative study of the slope of $\eta/s$ as a function of temperature, near the critical point, seems to be a promising avenue to explore. We leave some of these questions for the near future.

\section*{Acknowledgements}
M.B. acknowledges the support of the Shanghai Municipal Science and Technology Major Project (Grant No.2019SHZDZX01) and the sponsorship from the Yangyang Development Fund. 
S.C. acknowledges the support and hospitality of KITP and the Harvard University Physics Department, where parts of this work were carried on. 
 This research was supported in part by the National Science Foundation under Grant No. NSF PHY-1748958.
The work of S.C. was supported in part by the National Science Foundation under Grant No. PHY-2210271. 
The work of L.E. was supported in part by the NSF grant PHY-1915038. L.L. was partially supported by the National Natural Science Foundation of China Grants No.12122513, No.12075298 and No.12047503, and by the Chinese Academy of Sciences Project for Young Scientists in Basic Research YSBR-006.

\appendix

\section{Holographic Renormalization}\label{app:renorm}

In order to obtain a renormalized action, one should supplement (\ref{eq:action}) with appropriate boundary counterterms. The resulting action reads
\begin{align}\label{fullaction}
    S =& \int\!\mathrm{d}^5x\,\sqrt{-g} \, \left [ \frac{1}{2\kappa_5^2} \left( R + \frac{12}{L^2}\right) - \frac{1}{4\hat g^2} \, F^a_{MN} F^{aMN} \right]+ \int_{r=r_{bdy}}\!\mathrm{d}^4x\,\sqrt{-\gamma}\, \frac{1}{\kappa_5^2}K\notag\\
    &- \int_{r=r_{bdy}}\mathrm{d}^4x \, \sqrt{-\gamma}\frac{1}{\kappa_5^2}\left(\frac{3}{L}+\frac{1}{4}R[\gamma]+\frac{1}{8}R^{\mu\nu}[\gamma]R_{\mu\nu}[\gamma]\ln{r}-\frac{\alpha^2}{4}F^a_{\mu\nu}F^{a\mu\nu}\ln{r}\right)\,, 
\end{align}
with $r_{bdy}\rightarrow\infty$. 
Here $\gamma_{\mu\nu}=g_{\mu\nu}-n_\mu n_\nu$ is the induced metric,  $K_{\mu\nu}=\gamma^r{}_\mu\gamma^\sigma{}_\nu\nabla_r n_\sigma$ is the extrinsic curvature at the AdS boundary and $n^\mu$ is the outward pointing normal vector to the boundary. We will set $L=1$.

The expectation values of the energy-momentum tensor and current can then be obtained by varying the on-shell action~\eqref{fullaction}. We find
\begin{equation}\label{BYvev}
\begin{split}
    \langle T_{\mu\nu}\rangle=&\frac{1}{\kappa_5^2}\lim_{r\rightarrow\infty}r^2\Bigg(K\gamma_{\mu\nu}-K_{\mu\nu}-3\gamma_{\mu\nu}-{\alpha^2}\left(F^{a}_{\rho\mu}F^{a\rho}{}_{\nu}-\frac{1}{4}\gamma_{\mu\nu}F^a_{\rho\sigma}F^{a\rho\sigma}\right)\ln{r}+\frac{G_{\mu\nu}[\gamma]}{2}\\
    &+\frac{1}{2}G_\mu{}^\rho [\gamma]G_{\rho\nu}[\gamma]\ln{r}-\frac{1}{8}\left(G_{\rho\sigma} [\gamma]G^{\rho\sigma}[\gamma]+G^\sigma{}_\sigma[\gamma]-2\nabla_\sigma\nabla^\sigma R[\gamma]\right)\gamma_{\mu\nu}\ln{r}\\
    &-\frac{1}{4}\nabla_\mu\nabla_\nu R[\gamma]\ln{r}-\frac{1}{2}\nabla_\sigma\nabla_{(\mu} G_{\nu)}{}^\sigma\ln{r}+\frac{1}{2}G_{\mu\nu}[\gamma]R[\gamma]\ln{r}+\frac{1}{4}\nabla_\sigma\nabla^\sigma G_{\mu\nu}[\gamma]\ln{r}
    \Bigg),\\
  \langle J_a^{\mu} \rangle=&\frac{\alpha^2}{2\kappa_5^2}\lim_{r\rightarrow\infty}\sqrt{-\gamma}\left(-2n_\nu F^{a\mu\nu}+2\nabla_\nu F^{a\nu\mu}\ln{r}-2\epsilon^{abc}A^b_\nu F^{c\nu\mu}\ln{r}\right)\,,
\end{split}
\end{equation}
where $G_{\mu\nu}[\gamma]=R_{\mu\nu}[\gamma]-\frac{1}{2}R[\gamma]\gamma_{\mu\nu}$ and $R_{\mu\nu}[\gamma]$ is the Ricci tensor associated with the metric $\gamma_{\mu\nu}$.
In the coordinate system $\{t, r, x, y, z\}$ we used in our ansatz~\eqref{ansatz}, the metric including the shear perturbation is given by
\begin{align}
    g_{\mu\nu}=\begin{pmatrix}-u(r) & 0 & 0 & 0 & 0 \\
    0 & \frac{1}{u(r)} & 0 & 0 & 0\\
    0 & 0 & h(r) & e^{-i\omega t} v(r)\Psi_x(r) & 0 \\
    0 & 0 & e^{-i\omega t} v(r)\Psi_x(r)& v(r)  & 0\\
    0 & 0 & 0 & 0 & v(r)\end{pmatrix}\,,
\end{align}
and the $SU(2)$ gauge field reads
\begin{align}
    A_{1\mu}=\begin{pmatrix}0\\0\\w(r)\\e^{-i\omega t} a^1_y(r)\\0\end{pmatrix},\quad
    A_{2\mu}=\begin{pmatrix}0\\0\\0\\e^{-i\omega t} a^2_y(r)\\0\end{pmatrix},\quad
    A_{3_\mu}=\begin{pmatrix}\phi(r)\\0\\0\\0\\0\end{pmatrix}.
\end{align}
The boundary expansion for the background is given by

\begin{equation}\label{UVbk}
\begin{split}
    u(r)=&r ^2+\frac{u_{{b1}}-\frac{1}{3} \alpha ^2 \mu ^2 w_{{b0}}^2 \ln (r )}{r ^2}+\frac{u_{b2}}{r^4}+\frac{u_{b2l} \ln (r )}{r^4}+\frac{\alpha ^2\left(\frac{1}{30} \mu ^4 w_{{b0}}^2+\frac{1}{6} \mu ^2 w_{{b0}}^4\right)\ln ^2(r )}{r^4}+\ldots,\\
    v(r)=&r^2+\frac{\frac{1}{6} \alpha ^2 \mu ^2 w_{{b0}}^2 \ln (r )+v_{{b1}}}{r ^2}+\frac{v_{b2}}{r^4}+\frac{v_{b2l} \ln (r )}{r^4}+\frac{\frac{1}{30} \alpha ^2 \mu ^4 w_{{b0}}^2 \ln ^2(r )}{r^4}+\ldots,\\
    h(r)=&r^2+\frac{-\frac{1}{3} \alpha ^2 \mu ^2 w_{{b0}}^2 \ln (r )-\frac{1}{6} \alpha ^2 \mu ^2 w_{{b0}}^2-2 v_{{b1}}}{r ^2}+\frac{h_{b2}}{r^4}+\frac{h_{b2l} \ln (r )}{r^4}-\frac{2  \alpha ^2 \mu ^4 w_{{b0}}^2 \ln ^2(r )}{15 r^4}+\ldots,\\
    w(r)=& w_{{b0}}+\frac{\frac{1}{2} \mu ^2 w_{{b0}} \ln (r )+w_{{b1}}}{r ^2}+\frac{w_{b2}}{r ^4}+\frac{\ln (r ) \left(\frac{1}{8} \mu ^2 w_{{b0}}^3-\frac{\mu ^4 w_{{b0}}}{16}\right)}{r^4}+\ldots.\\
    \phi(r)=&\mu+\frac{\phi _{{b1}}-\frac{1}{2} \mu  w_{{b0}}^2 \ln (r )}{r ^2}+\frac{\phi_{b2}}{r^4}+\frac{\ln (r ) \left(\frac{1}{8} \mu ^3 w_{{b0}}^2-\frac{\mu  w_{{b0}}^4}{16}\right)}{r^4}+\ldots,
\end{split}
\end{equation}
where
\begin{equation}
\begin{split}
    u_{b2}&=\alpha ^2 \left(\frac{17}{75} \mu ^2 w_{{b0}} w_{{b1}}+\frac{1}{15} \mu  w_{{b0}}^2 \phi _{{b1}}+\frac{81 \mu ^4 w_{{b0}}^2}{1000}-\frac{3}{200} \mu ^2 w_{{b0}}^4+\frac{2 w_{{b1}}^2}{15}+\frac{2 \phi _{{b1}}^2}{3}\right)\, ,\\
    u_{b2l}&=\alpha^2\left(\frac{2}{15} \mu ^2 w_{{b0}} w_{{b1}}-\frac{2}{3} \mu  w_{{b0}}^2 \phi _{{b1}}+\frac{17}{150} \mu ^4 w_{{b0}}^2-\frac{1}{30} \mu ^2 w_{{b0}}^4\right)\, ,\\
    v_{b2}&=\alpha ^2 \left(-\frac{8}{75} \mu ^2 w_{{b0}} w_{{b1}}-\frac{7}{45} \mu  w_{{b0}}^2 \phi _{{b1}}-\frac{271 \mu ^4 w_{{b0}}^2}{9000}+\frac{61 \mu ^2 w_{{b0}}^4}{1350}+\frac{2 w_{{b1}}^2}{15}\right)\, ,\\
    v_{b2l}&=\alpha ^2\left(\frac{2}{15} \mu ^2 w_{{b0}} w_{{b1}}+\frac{1}{75} (-4) \mu ^4 w_{{b0}}^2+\frac{7}{90} \mu ^2 w_{{b0}}^4\right)\, ,\\
    h_{b2}&=\alpha ^2 \left(\frac{26}{225} \mu ^2 w_{{b0}} w_{{b1}}+\frac{8}{45} \mu  w_{{b0}}^2 \phi _{{b1}}+\frac{203 \mu ^4 w_{{b0}}^2}{6750}-\frac{89 \mu ^2 w_{{b0}}^4}{1350}-\frac{8 w_{{b1}}^2}{15}\right)\, ,\\
    h_{b2l}&=\alpha ^2 \left(-\frac{8}{15} \mu ^2 w_{{b0}} w_{{b1}}+\frac{13}{225} \mu ^4 w_{{b0}}^2-\frac{4}{45} \mu ^2 w_{{b0}}^4\right)\, ,\\
    w_{b2}&=-\frac{3}{64} \mu ^4 w_{{b0}}+\frac{3}{32} \mu ^2 w_{{b0}}^3-\frac{1}{4} \mu  w_{{b0}} \phi _{{b1}}-\frac{1}{8} \mu ^2 w_{{b1}}\, ,\\
    \phi_{b2}&=\frac{3}{32} \mu ^3 w_{{b0}}^2-\frac{3 \mu  w_{{b0}}^4}{64}+\frac{1}{4} \mu  w_{{b0}} w_{{b1}}+\frac{1}{8} w_{{b0}}^2 \phi _{{b1}}\,.
\end{split}
\end{equation}
In the expansions above, we have taken the normalization of the time
coordinate at the boundary such that $u(r\rightarrow \infty)=1$. Also, $w_{b0}$ is the source that breaks the rotational symmetry explicitly.
The boundary expansion for the perturbations reads
\begin{equation}\label{UVperturb}
\begin{split}
\Psi_x=&(\Psi_x)^b_0+\frac{\omega^2(\Psi_x)^b_0}{4r^2}+\frac{(\Psi_x)^b_2}{r^4}
    +\frac{-8\alpha^2\mu^2w_{b0}(a^1_y)^b_0+8i\omega\alpha^2\mu w_{b0}(a^2_y)^b_0+\omega^4(\Psi_x)^b_0}{16r^4}\ln{r}+\ldots\,,\\
    a^1_y=&(a^1_y)^b_0+\frac{(a^1_y)^b_1}{r^2}-\frac{2i\omega\mu(a^2_y)^b_0-(a^1_y)^b_0(\mu^2+\omega^2)}{2 r^2}\ln{r}+\ldots\,,\\
    a^2_y=&(a^2_y)^b_0+\frac{(a^2_y)^b_1}{r^2}-\frac{(w_{b0}^2-\mu^2-\omega^2)(a^2_y)^b_0-2i\omega\mu(a^1_y)^b_0+i\omega\mu w_{b0}(\Psi_x)^b_0}{2 r^2}\ln{r}+\ldots.
\end{split}
\end{equation}

Substituting the expansions~\eqref{UVbk} and~\eqref{UVperturb} into~\eqref{BYvev}, 
we obtain
\begin{equation}
\begin{split}
    \langle T_{tt\rangle}&=-\frac{9 u_{b1}+2 \alpha^2 w_{b0}^2\mu^2}{6\kappa_5^2}\,,\\
    \langle T_{xx}\rangle&=-\frac{u_{b1}+8v_{b1}}{2\kappa_5^2}\,,\\
    \langle T_{yy}\rangle&=\langle T_{zz}\rangle=\frac{\alpha^2  \mu ^2 w_{b0}^2-6 u_{b1}+24 v_{b1}}{12 \kappa _5^2}\,,\\
    \langle J_1^x \rangle&=\frac{\alpha^2  \left(4 w_{b1}-\mu ^2 w_{b0}\right)}{2 \kappa _5^2}\,,\\
    \langle J^t_3\rangle&=-\frac{\alpha^2  \left(\mu  w_{{b0}}^2+4 \phi _{{b1}}\right)}{2 \kappa _5^2}\,,
\end{split}
\end{equation}
and
\begin{align}
    \langle T_{xy} \rangle=& \langle T_{yy}\rangle(\Psi_x)^b_0+\frac{1}{12\kappa_5^2}\left[ 3 \alpha^2  \mu ^2   w_{{b0}}(a^1_y)^b_0 -3 i \alpha^2  \mu     \omega  w_{{b0}}(a^2_y)^b_0+24 (\Psi_x)^b_2 \right],\\
    \langle J^y_1\rangle=&\frac{\alpha^2}{2\kappa_5^2}\left[4(a^1_y)^b_1-\mu^2(a^1_y)^b_0+(w_{b0}\mu^2-4w_{b1})(\Psi_x)^b_0+2i\omega\mu(a^2_y)^b_0\right],\\
    \langle J^y_2 \rangle=&\frac{\alpha^2}{2\kappa_5^2}\left[4(a^2_y)^b_1+(w_{b0}^2-\mu^2)(a^2_y)^b_0-2i\omega\mu (a^1_y)^b_0+i\omega\mu w_{b0}(\Psi_x)^b_0\right]\,,
\end{align}
with all other components vanishing. 
Importantly, the terms $(a^1_y)^b_1$ and $(a^2_y)^b_1$ in the expressions above correspond to the sources for the fluctuations of the $J_{y1}$ and $J_{y2}$ current operators on the boundary, respectively. In order to compute the shear viscosity, we turn off such terms and obtain\footnote{Terms proportional to $(a^1_y)^b_1$ and $(a^2_y)^b_1$ would contribute to mixed Green's functions involving the stress tensor and the currents $J^y_1$ and $J^y_2$ .} 
\begin{equation}
 \langle T_{xy} \rangle=\left[ \langle T_{yy} \rangle+\frac{2}{\kappa^2_5}\frac{(\Psi_x)^b_2}{(\Psi_x)^b_0} \right] (\Psi_x)^b_0\,.
\end{equation}
In a transversely isotropic fluid, one has
\begin{equation}\label{etaby}
 \langle T_{xy} \rangle=\left[ \langle T_{yy} \rangle+i\omega \eta \right] (\Psi_x)^b_0\,,
\end{equation}
in the low frequency limit. Then one obtains
\begin{equation}
    \eta=-\frac{2}{\kappa^2_5}\lim_{\omega\to 0}\frac{i}{\omega}\frac{(\Psi_x)^b_2}{(\Psi_x)^b_0}\,,
\end{equation}
which is consistent with the well-known Kubo's formula~\eqref{eqkubo}.

\section{Horizon formula for $\eta/s$}\label{two}
We relegate to this section some of the details of the derivation of Eq.\eqref{res}
which are not included in the main text. 
We begin with Eq.\eqref{fullhxy} for the redefined shear mode $h_{xy}=v(r)\Psi_x(r)$. We expand its solution in a perturbative series with respect to the frequency $\omega$
\begin{equation}
    \Psi_x(r)=u(r)^{-i\omega/(4\pi T)}\left(1+\omega\,\Psi_x^{(1)}(r)+\dots
    \right)\,.\
\end{equation} 
where higher order terms are ignored, since irrelevant for the computation of the zero frequency viscosity, and the solution is forced to obey ingoing boundary conditions at the horizon $r=r_h$. The leading order solution is given by
\begin{equation}
\label{finalshear}
     \Psi_x^{(1)}(r)=\int_{r_h}^r \left[ \frac{i}{4\pi T}\frac{u'}{u}-\frac{2 \alpha^2 a_{y1}^{(1)} w'}{v}-\frac{i v_1^2}{\sqrt{h_1}} \frac{\sqrt{h}}{uv^2}\right] d \tilde{r}\,. 
    \end{equation}
According to ~\eqref{etaby}, in order to compute $\eta$, we need to know the leading and subleading coefficients of the boundary expansion of $\Psi_x$ in Eq.\eqref{UVperturb}, \emph{i.e.} $(\Psi_x)^b_0$ and $(\Psi_x)^b_2$.

To compute the integral of $\Psi_x^{(1)}$, we consider the following coordinate transformation $z=\frac{1}{r^2}$ and  $\tilde{z}=\frac{1}{\tilde{r}^2}$. Then $\Psi_x^{(1)}$ becomes
\begin{align}
   \Psi_x^{(1)}=\int_{z_h}^z \left[ \frac{i}{4\pi T}\frac{u'}{u}-\frac{2 \alpha^2 a_{y1}^{(1)} w'}{v}+\frac{i v_1^2}{\sqrt{h_1}} \frac{\sqrt{h}}{2\tilde{z}^{3/2}uv^2}\right] d\tilde{z}\,,
\end{align}
where the prime appearing in the integrand denotes the derivative with respect to $\tilde{z}$, and $z_h=1/r_h^2$. Note that the AdS boundary now corresponds to $z=0$. This integral can be decomposed into several parts
\begin{align}
    \Psi_x^{(1)}=&\int_{z_h}^z \left[ \frac{i}{4\pi T}\left(\frac{u'}{u}+\frac{1}{\tilde{z}}\right)-\frac{i}{4\pi T}\frac{1}{\tilde{z}}-\frac{2 \alpha^2 a_{y1}^{(1)} w'}{v}+\frac{i v_1^2}{\sqrt{h_1}} \frac{\sqrt{h}}{2\tilde{z}^{3/2}uv^2}\right] d\tilde{z}\,,\notag\\
    =&C-\int_{z_h}^z\frac{i}{4\pi T\tilde{z}}d\tilde{z}+\int_0^z \frac{i}{4\pi T} \left(\frac{u'}{u}+\frac{1}{\tilde{z}}\right)d\tilde{z}+\int_0^z\left(-\frac{2 \alpha^2 a_{y1}^{(1)} w'}{v}+\frac{i v_1^2}{\sqrt{h_1}} \frac{\sqrt{h}}{2\tilde{z}^{3/2}uv^2}\right)d\tilde{z}\,,
\end{align}
where $C$ is a constant given by
\begin{align}
    C=\int_{z_h}^0 \left[\frac{i}{4\pi T} \left(\frac{u'}{u}+\frac{1}{\tilde{z}}\right)-\frac{2 \alpha^2 a_{y1}^{(1)} w'}{v}+\frac{i v_1^2}{\sqrt{h_1}} \frac{\sqrt{h}}{2\tilde{z}^{3/2}uv^2}\right] d\tilde{z}.
\end{align}

We set to zero the source 
for the gauge field perturbation, since it is not relevant for the $\eta/s$ calculation. Then, the UV expansion of $a^{(1)}_{y1}$ reads
\begin{equation}
a^{(1)}_{y1}= (a^{(1)}_{y1})_1^b\, z+\ldots
\end{equation}
in terms of the new radial coordinate $z$. Substituting the boundary expansion~\eqref{UVbk}, we find
\begin{align}
    \Psi_x^{(1)}=C-\frac{i }{4 \pi  T}\ln \left(\frac{z}{z_h}\right)+\frac{i {v_1}^2 z^2}{4 \sqrt{{h_1}}}+\frac{i z^2 \left(6{u_{b1}}+{\alpha^2} \mu ^2 {w_{b0}}^2 \ln z\right)}{24 \pi T}+\ldots
\end{align}
near the AdS boundary $z=0$. Finally, we obtain the boundary expansion of $\Psi_x$:
\begin{align}\label{UVPsi}
    \Psi_x&=u(r)^{-i\omega/(4\pi T)}\left(1+\omega\,\Psi_x^{(1)}
    \right)\,
    =1+\omega \left(C+\frac{i\ln{z_h}}{4\pi T}\right)+i\omega\frac{v_1^2z^2}{4\sqrt{h_1}}+\dots\,,
\end{align}
where we have made use of the approximation
\begin{equation}
    u(r)^{-i\omega/(4\pi T)} \sim 1 - \frac{i\omega}{4 \pi T}\ln{u} \, ,
\end{equation}
valid in the small $\omega/T$ regime.

Therefore, we find from~\eqref{UVPsi} that
\begin{equation}
 (\Psi_x)^b_0= 1+\omega \left(C+\frac{i\ln{z_h}}{4\pi T}\right),\quad (\Psi_x)^b_2= i\omega\frac{v_1^2}{4\sqrt{h_1}}\,, 
\end{equation}
to linear order of frequency. Then, comparing the above equation with linear response theory, Eq.\eqref{ok}, or equivalently using~\eqref{etaby}, we obtain the shear viscosity
\begin{equation}
    \eta=-\frac{2}{\kappa^2_5}\lim_{\omega\to 0}\frac{i}{\omega}\frac{(\Psi_x)^b_2}{(\Psi_x)^b_0}=\frac{1}{2\kappa_5^2}\frac{ v_1^2}{\sqrt{h_1}}\,,
\end{equation}
which is entirely determined by the horizon data. Finally, 
using $s=\frac{2\pi}{\kappa_5^2}\sqrt{h_1}v_1$,
the ratio of shear viscosity over entropy density is given by
\begin{equation}
    \frac{\eta}{s}=\frac{1}{4\pi}\frac{v_1}{h_1}\,,
\end{equation}
as reported in the main text.

\bibliographystyle{JHEP}
\bibliography{aniso}

\providecommand{\href}[2]{#2}\begingroup\raggedright\begin{thebibliography}{10}

\bibitem{Policastro:2001yc}
G.~Policastro, D.~T. Son and A.~O. Starinets, \emph{{The Shear viscosity of
  strongly coupled N=4 supersymmetric Yang-Mills plasma}},
  \href{http://dx.doi.org/10.1103/PhysRevLett.87.081601}{\emph{Phys. Rev.
  Lett.} {\bf 87} (2001) 081601},
  [\href{http://arxiv.org/abs/hep-th/0104066}{{\tt hep-th/0104066}}].

\bibitem{Buchel:2003tz}
A.~Buchel and J.~T. Liu, \emph{{Universality of the shear viscosity in
  supergravity}},
  \href{http://dx.doi.org/10.1103/PhysRevLett.93.090602}{\emph{Phys. Rev.
  Lett.} {\bf 93} (2004) 090602},
  [\href{http://arxiv.org/abs/hep-th/0311175}{{\tt hep-th/0311175}}].

\bibitem{Kovtun:2003wp}
P.~Kovtun, D.~T. Son and A.~O. Starinets, \emph{{Holography and hydrodynamics:
  Diffusion on stretched horizons}},
  \href{http://dx.doi.org/10.1088/1126-6708/2003/10/064}{\emph{JHEP} {\bf 10}
  (2003) 064}, [\href{http://arxiv.org/abs/hep-th/0309213}{{\tt
  hep-th/0309213}}].

\bibitem{Kovtun:2004de}
P.~Kovtun, D.~T. Son and A.~O. Starinets, \emph{{Viscosity in strongly
  interacting quantum field theories from black hole physics}},
  \href{http://dx.doi.org/10.1103/PhysRevLett.94.111601}{\emph{Phys. Rev.
  Lett.} {\bf 94} (2005) 111601},
  [\href{http://arxiv.org/abs/hep-th/0405231}{{\tt hep-th/0405231}}].

\bibitem{Cremonini:2011iq}
S.~Cremonini, \emph{{The Shear Viscosity to Entropy Ratio: A Status Report}},
  \href{http://dx.doi.org/10.1142/S0217984911027315}{\emph{Mod. Phys. Lett. B}
  {\bf 25} (2011) 1867--1888}, [\href{http://arxiv.org/abs/1108.0677}{{\tt
  1108.0677}}].

\bibitem{PhysRevLett.99.021602}
T.~D. Cohen, \emph{Is there a ``most perfect fluid'' consistent with quantum
  field theory?},
  \href{http://dx.doi.org/10.1103/PhysRevLett.99.021602}{\emph{Phys. Rev.
  Lett.} {\bf 99} (Jul, 2007) 021602}.

\bibitem{Cherman:2007fj}
A.~Cherman, T.~D. Cohen and P.~M. Hohler, \emph{{A Sticky business: The Status
  of the cojectured viscosity/entropy density bound}},
  \href{http://dx.doi.org/10.1088/1126-6708/2008/02/026}{\emph{JHEP} {\bf 02}
  (2008) 026}, [\href{http://arxiv.org/abs/0708.4201}{{\tt 0708.4201}}].

\bibitem{Dobado:2007tm}
A.~Dobado and F.~J. Llanes-Estrada, \emph{{On the violation of the holographic
  viscosity versus entropy KSS bound in non relativistic systems}},
  \href{http://dx.doi.org/10.1140/epjc/s10052-007-0332-5}{\emph{Eur. Phys. J.
  C} {\bf 51} (2007) 913--918},
  [\href{http://arxiv.org/abs/hep-th/0703132}{{\tt hep-th/0703132}}].

\bibitem{PhysRevLett.100.029101}
D.~T. Son, \emph{Comment on ``is there a `most perfect fluid' consistent with
  quantum field theory?''},
  \href{http://dx.doi.org/10.1103/PhysRevLett.100.029101}{\emph{Phys. Rev.
  Lett.} {\bf 100} (Jan, 2008) 029101}.

\bibitem{doi:10.1126/sciadv.aba3747}
K.~Trachenko and V.~V. Brazhkin, \emph{Minimal quantum viscosity from
  fundamental physical constants},
  \href{http://dx.doi.org/10.1126/sciadv.aba3747}{\emph{Science Advances} {\bf
  6} (2020) eaba3747},
  [\href{http://arxiv.org/abs/https://www.science.org/doi/pdf/10.1126/sciadv.aba3747}{{\tt
  https://www.science.org/doi/pdf/10.1126/sciadv.aba3747}}].

\bibitem{PhysRevB.103.014311}
K.~Trachenko, M.~Baggioli, K.~Behnia and V.~V. Brazhkin, \emph{Universal lower
  bounds on energy and momentum diffusion in liquids},
  \href{http://dx.doi.org/10.1103/PhysRevB.103.014311}{\emph{Phys. Rev. B} {\bf
  103} (Jan, 2021) 014311}.

\bibitem{Buchel:2004di}
A.~Buchel, J.~T. Liu and A.~O. Starinets, \emph{{Coupling constant dependence
  of the shear viscosity in N=4 supersymmetric Yang-Mills theory}},
  \href{http://dx.doi.org/10.1016/j.nuclphysb.2004.11.055}{\emph{Nucl. Phys. B}
  {\bf 707} (2005) 56--68}, [\href{http://arxiv.org/abs/hep-th/0406264}{{\tt
  hep-th/0406264}}].

\bibitem{Buchel:2008vz}
A.~Buchel, R.~C. Myers and A.~Sinha, \emph{{Beyond eta/s = 1/4 pi}},
  \href{http://dx.doi.org/10.1088/1126-6708/2009/03/084}{\emph{JHEP} {\bf 03}
  (2009) 084}, [\href{http://arxiv.org/abs/0812.2521}{{\tt 0812.2521}}].

\bibitem{Kats:2007mq}
Y.~Kats and P.~Petrov, \emph{{Effect of curvature squared corrections in AdS on
  the viscosity of the dual gauge theory}},
  \href{http://dx.doi.org/10.1088/1126-6708/2009/01/044}{\emph{JHEP} {\bf 01}
  (2009) 044}, [\href{http://arxiv.org/abs/0712.0743}{{\tt 0712.0743}}].

\bibitem{Myers:2008yi}
R.~C. Myers, M.~F. Paulos and A.~Sinha, \emph{{Quantum corrections to eta/s}},
  \href{http://dx.doi.org/10.1103/PhysRevD.79.041901}{\emph{Phys. Rev. D} {\bf
  79} (2009) 041901}, [\href{http://arxiv.org/abs/0806.2156}{{\tt 0806.2156}}].

\bibitem{Brigante:2007nu}
M.~Brigante, H.~Liu, R.~C. Myers, S.~Shenker and S.~Yaida, \emph{{Viscosity
  Bound Violation in Higher Derivative Gravity}},
  \href{http://dx.doi.org/10.1103/PhysRevD.77.126006}{\emph{Phys. Rev. D} {\bf
  77} (2008) 126006}, [\href{http://arxiv.org/abs/0712.0805}{{\tt 0712.0805}}].

\bibitem{Brigante:2008gz}
M.~Brigante, H.~Liu, R.~C. Myers, S.~Shenker and S.~Yaida, \emph{{The Viscosity
  Bound and Causality Violation}},
  \href{http://dx.doi.org/10.1103/PhysRevLett.100.191601}{\emph{Phys. Rev.
  Lett.} {\bf 100} (2008) 191601}, [\href{http://arxiv.org/abs/0802.3318}{{\tt
  0802.3318}}].

\bibitem{Buchel:2010wf}
A.~Buchel and S.~Cremonini, \emph{{Viscosity Bound and Causality in Superfluid
  Plasma}}, \href{http://dx.doi.org/10.1007/JHEP10(2010)026}{\emph{JHEP} {\bf
  10} (2010) 026}, [\href{http://arxiv.org/abs/1007.2963}{{\tt 1007.2963}}].

\bibitem{Ling:2016ien}
Y.~Ling, Z.-Y. Xian and Z.~Zhou, \emph{{Holographic Shear Viscosity in
  Hyperscaling Violating Theories without Translational Invariance}},
  \href{http://dx.doi.org/10.1007/JHEP11(2016)007}{\emph{JHEP} {\bf 11} (2016)
  007}, [\href{http://arxiv.org/abs/1605.03879}{{\tt 1605.03879}}].

\bibitem{Hartnoll:2016tri}
S.~A. Hartnoll, D.~M. Ramirez and J.~E. Santos, \emph{{Entropy production,
  viscosity bounds and bumpy black holes}},
  \href{http://dx.doi.org/10.1007/JHEP03(2016)170}{\emph{JHEP} {\bf 03} (2016)
  170}, [\href{http://arxiv.org/abs/1601.02757}{{\tt 1601.02757}}].

\bibitem{Alberte:2016xja}
L.~Alberte, M.~Baggioli and O.~Pujolas, \emph{{Viscosity bound violation in
  holographic solids and the viscoelastic response}},
  \href{http://dx.doi.org/10.1007/JHEP07(2016)074}{\emph{JHEP} {\bf 07} (2016)
  074}, [\href{http://arxiv.org/abs/1601.03384}{{\tt 1601.03384}}].

\bibitem{Burikham:2016roo}
P.~Burikham and N.~Poovuttikul, \emph{{Shear viscosity in holography and
  effective theory of transport without translational symmetry}},
  \href{http://dx.doi.org/10.1103/PhysRevD.94.106001}{\emph{Phys. Rev. D} {\bf
  94} (2016) 106001}, [\href{http://arxiv.org/abs/1601.04624}{{\tt
  1601.04624}}].

\bibitem{Baggioli:2020ljz}
M.~Baggioli and W.-J. Li, \emph{{Universal Bounds on Transport in Holographic
  Systems with Broken Translations}},
  \href{http://arxiv.org/abs/2005.06482}{{\tt 2005.06482}}.

\bibitem{Baggioli:2022aft}
M.~Baggioli and G.~Frangi, \emph{{Holographic supersolids}},
  \href{http://dx.doi.org/10.1007/JHEP06(2022)152}{\emph{JHEP} {\bf 06} (2022)
  152}, [\href{http://arxiv.org/abs/2202.03745}{{\tt 2202.03745}}].

\bibitem{RevModPhys.95.011001}
M.~Baggioli and B.~Gout\'eraux, \emph{Colloquium: Hydrodynamics and holography
  of charge density wave phases},
  \href{http://dx.doi.org/10.1103/RevModPhys.95.011001}{\emph{Rev. Mod. Phys.}
  {\bf 95} (Jan, 2023) 011001}.

\bibitem{Landsteiner:2007bd}
K.~Landsteiner and J.~Mas, \emph{{The Shear viscosity of the non-commutative
  plasma}}, \href{http://dx.doi.org/10.1088/1126-6708/2007/07/088}{\emph{JHEP}
  {\bf 07} (2007) 088}, [\href{http://arxiv.org/abs/0706.0411}{{\tt
  0706.0411}}].

\bibitem{Liu:2016njg}
H.-S. Liu, H.~Lu and C.~N. Pope, \emph{{Magnetically-Charged Black Branes and
  Viscosity/Entropy Ratios}},
  \href{http://dx.doi.org/10.1007/JHEP12(2016)097}{\emph{JHEP} {\bf 12} (2016)
  097}, [\href{http://arxiv.org/abs/1602.07712}{{\tt 1602.07712}}].

\bibitem{Giataganas:2017koz}
D.~Giataganas, U.~G\"ursoy and J.~F. Pedraza, \emph{{Strongly-coupled
  anisotropic gauge theories and holography}},
  \href{http://dx.doi.org/10.1103/PhysRevLett.121.121601}{\emph{Phys. Rev.
  Lett.} {\bf 121} (2018) 121601}, [\href{http://arxiv.org/abs/1708.05691}{{\tt
  1708.05691}}].

\bibitem{Donos:2016zpf}
A.~Donos, J.~P. Gauntlett and O.~Sosa-Rodriguez, \emph{{Anisotropic plasmas
  from axion and dilaton deformations}},
  \href{http://dx.doi.org/10.1007/JHEP11(2016)002}{\emph{JHEP} {\bf 11} (2016)
  002}, [\href{http://arxiv.org/abs/1608.02970}{{\tt 1608.02970}}].

\bibitem{Ge:2015owa}
X.-H. Ge, \emph{{Notes on shear viscosity bound violation in anisotropic
  models}}, \href{http://dx.doi.org/10.1007/s11433-015-5776-2}{\emph{Sci. China
  Phys. Mech. Astron.} {\bf 59} (2016) 630401},
  [\href{http://arxiv.org/abs/1510.06861}{{\tt 1510.06861}}].

\bibitem{Giataganas:2013lga}
D.~Giataganas, \emph{{Observables in Strongly Coupled Anisotropic Theories}},
  \href{http://dx.doi.org/10.22323/1.177.0122}{\emph{PoS} {\bf Corfu2012}
  (2013) 122}, [\href{http://arxiv.org/abs/1306.1404}{{\tt 1306.1404}}].

\bibitem{Chakraborty:2017msh}
S.~Chakraborty and R.~Samanta, \emph{{Viscosity for anisotropic
  Reissner-Nordstr\"om black branes}},
  \href{http://dx.doi.org/10.1103/PhysRevD.95.106012}{\emph{Phys. Rev. D} {\bf
  95} (2017) 106012}, [\href{http://arxiv.org/abs/1702.07874}{{\tt
  1702.07874}}].

\bibitem{Mateos:2011ix}
D.~Mateos and D.~Trancanelli, \emph{{The anisotropic N=4 super Yang-Mills
  plasma and its instabilities}},
  \href{http://dx.doi.org/10.1103/PhysRevLett.107.101601}{\emph{Phys. Rev.
  Lett.} {\bf 107} (2011) 101601}, [\href{http://arxiv.org/abs/1105.3472}{{\tt
  1105.3472}}].

\bibitem{Rebhan:2011vd}
A.~Rebhan and D.~Steineder, \emph{{Violation of the Holographic Viscosity Bound
  in a Strongly Coupled Anisotropic Plasma}},
  \href{http://dx.doi.org/10.1103/PhysRevLett.108.021601}{\emph{Phys. Rev.
  Lett.} {\bf 108} (2012) 021601}, [\href{http://arxiv.org/abs/1110.6825}{{\tt
  1110.6825}}].

\bibitem{Jain:2014vka}
S.~Jain, N.~Kundu, K.~Sen, A.~Sinha and S.~P. Trivedi, \emph{{A Strongly
  Coupled Anisotropic Fluid From Dilaton Driven Holography}},
  \href{http://dx.doi.org/10.1007/JHEP01(2015)005}{\emph{JHEP} {\bf 01} (2015)
  005}, [\href{http://arxiv.org/abs/1406.4874}{{\tt 1406.4874}}].

\bibitem{Jain:2015txa}
S.~Jain, R.~Samanta and S.~P. Trivedi, \emph{{The Shear Viscosity in
  Anisotropic Phases}},
  \href{http://dx.doi.org/10.1007/JHEP10(2015)028}{\emph{JHEP} {\bf 10} (2015)
  028}, [\href{http://arxiv.org/abs/1506.01899}{{\tt 1506.01899}}].

\bibitem{Azeyanagi:2009pr}
T.~Azeyanagi, W.~Li and T.~Takayanagi, \emph{{On String Theory Duals of
  Lifshitz-like Fixed Points}},
  \href{http://dx.doi.org/10.1088/1126-6708/2009/06/084}{\emph{JHEP} {\bf 06}
  (2009) 084}, [\href{http://arxiv.org/abs/0905.0688}{{\tt 0905.0688}}].

\bibitem{Rath:2020beo}
S.~Rath and B.~K. Patra, \emph{{Viscous properties of hot and dense QCD matter
  in the presence of a magnetic field}},
  \href{http://dx.doi.org/10.1140/epjc/s10052-021-08931-1}{\emph{Eur. Phys. J.
  C} {\bf 81} (2021) 139}, [\href{http://arxiv.org/abs/2010.02886}{{\tt
  2010.02886}}].

\bibitem{Gursoy:2020kjd}
U.~G\"ursoy, M.~J\"arvinen, G.~Nijs and J.~F. Pedraza, \emph{{On the interplay
  between magnetic field and anisotropy in holographic QCD}},
  \href{http://dx.doi.org/10.1007/JHEP03(2021)180}{\emph{JHEP} {\bf 03} (2021)
  180}, [\href{http://arxiv.org/abs/2011.09474}{{\tt 2011.09474}}].

\bibitem{Finazzo:2016mhm}
S.~I. Finazzo, R.~Critelli, R.~Rougemont and J.~Noronha, \emph{{Momentum
  transport in strongly coupled anisotropic plasmas in the presence of strong
  magnetic fields}},
  \href{http://dx.doi.org/10.1103/PhysRevD.94.054020}{\emph{Phys. Rev. D} {\bf
  94} (2016) 054020}, [\href{http://arxiv.org/abs/1605.06061}{{\tt
  1605.06061}}].

\bibitem{Critelli:2014kra}
R.~Critelli, S.~I. Finazzo, M.~Zaniboni and J.~Noronha, \emph{{Anisotropic
  shear viscosity of a strongly coupled non-Abelian plasma from magnetic
  branes}}, \href{http://dx.doi.org/10.1103/PhysRevD.90.066006}{\emph{Phys.
  Rev. D} {\bf 90} (2014) 066006}, [\href{http://arxiv.org/abs/1406.6019}{{\tt
  1406.6019}}].

\bibitem{Ammon:2012qs}
M.~Ammon, V.~G. Filev, J.~Tarrio and D.~Zoakos, \emph{{D3/D7 Quark-Gluon Plasma
  with Magnetically Induced Anisotropy}},
  \href{http://dx.doi.org/10.1007/JHEP09(2012)039}{\emph{JHEP} {\bf 09} (2012)
  039}, [\href{http://arxiv.org/abs/1207.1047}{{\tt 1207.1047}}].

\bibitem{Natsuume:2010ky}
M.~Natsuume and M.~Ohta, \emph{{The Shear viscosity of holographic
  superfluids}}, \href{http://dx.doi.org/10.1143/PTP.124.931}{\emph{Prog.
  Theor. Phys.} {\bf 124} (2010) 931--951},
  [\href{http://arxiv.org/abs/1008.4142}{{\tt 1008.4142}}].

\bibitem{Basu:2011tt}
P.~Basu and J.-H. Oh, \emph{{Analytic Approaches to Anisotropic Holographic
  Superfluids}}, \href{http://dx.doi.org/10.1007/JHEP07(2012)106}{\emph{JHEP}
  {\bf 07} (2012) 106}, [\href{http://arxiv.org/abs/1109.4592}{{\tt
  1109.4592}}].

\bibitem{Erdmenger:2011tj}
J.~Erdmenger, P.~Kerner and H.~Zeller, \emph{{Transport in Anisotropic
  Superfluids: A Holographic Description}},
  \href{http://dx.doi.org/10.1007/JHEP01(2012)059}{\emph{JHEP} {\bf 01} (2012)
  059}, [\href{http://arxiv.org/abs/1110.0007}{{\tt 1110.0007}}].

\bibitem{Bhattacharyya:2014wfa}
A.~Bhattacharyya and D.~Roychowdhury, \emph{{Viscosity bound for anisotropic
  superfluids in higher derivative gravity}},
  \href{http://dx.doi.org/10.1007/JHEP03(2015)063}{\emph{JHEP} {\bf 03} (2015)
  063}, [\href{http://arxiv.org/abs/1410.3222}{{\tt 1410.3222}}].

\bibitem{Oh:2012zu}
J.-H. Oh, \emph{{Running Shear Viscosities in An-Isotropic Holographic
  Superfluids}}, \href{http://dx.doi.org/10.1007/JHEP06(2012)103}{\emph{JHEP}
  {\bf 06} (2012) 103}, [\href{http://arxiv.org/abs/1201.5605}{{\tt
  1201.5605}}].

\bibitem{Landsteiner:2016stv}
K.~Landsteiner, Y.~Liu and Y.-W. Sun, \emph{{Odd viscosity in the quantum
  critical region of a holographic Weyl semimetal}},
  \href{http://dx.doi.org/10.1103/PhysRevLett.117.081604}{\emph{Phys. Rev.
  Lett.} {\bf 117} (2016) 081604}, [\href{http://arxiv.org/abs/1604.01346}{{\tt
  1604.01346}}].

\bibitem{Moradpouri:2022zwa}
A.~Moradpouri, S.~A. Jafari and M.~Torabian, \emph{{Holographic Hydrodynamics
  of $Tilted$ Dirac Materials}},  \href{http://arxiv.org/abs/2211.15289}{{\tt
  2211.15289}}.

\bibitem{Polchinski:2012nh}
J.~Polchinski and E.~Silverstein, \emph{{Large-density field theory, viscosity,
  and '$2k_F$' singularities from string duals}},
  \href{http://dx.doi.org/10.1088/0264-9381/29/19/194008}{\emph{Class. Quant.
  Grav.} {\bf 29} (2012) 194008}, [\href{http://arxiv.org/abs/1203.1015}{{\tt
  1203.1015}}].

\bibitem{Penin:2017lqt}
J.~M. Penin, A.~V. Ramallo and D.~Zoakos, \emph{{Anisotropic D3-D5 black holes
  with unquenched flavors}},
  \href{http://dx.doi.org/10.1007/JHEP02(2018)139}{\emph{JHEP} {\bf 02} (2018)
  139}, [\href{http://arxiv.org/abs/1710.00548}{{\tt 1710.00548}}].

\bibitem{Samanta:2016pic}
R.~Samanta, R.~Sharma and S.~P. Trivedi, \emph{{Shear viscosity in an
  anisotropic unitary Fermi gas}},
  \href{http://dx.doi.org/10.1103/PhysRevA.96.053601}{\emph{Phys. Rev. A} {\bf
  96} (2017) 053601}, [\href{http://arxiv.org/abs/1607.04799}{{\tt
  1607.04799}}].

\bibitem{kim2021hydrodynamic}
S.~W. Kim, G.~Jose and B.~Uchoa, \emph{Hydrodynamic transport and violation of
  the viscosity-to-entropy ratio bound in nodal-line semimetals},
  {\emph{Physical Review Research} {\bf 3} (2021) 033003}.

\bibitem{gochan2019viscosity}
M.~P. Gochan, H.~Li and K.~S. Bedell, \emph{Viscosity bound violation in
  viscoelastic fermi liquids}, {\emph{Journal of Physics Communications} {\bf
  3} (2019) 065008}.

\bibitem{Link:2017ora}
J.~M. Link, B.~N. Narozhny, E.~I. Kiselev and J.~Schmalian,
  \emph{{Out-of-bounds hydrodynamics in anisotropic Dirac fluids}},
  \href{http://dx.doi.org/10.1103/PhysRevLett.120.196801}{\emph{Phys. Rev.
  Lett.} {\bf 120} (2018) 196801}, [\href{http://arxiv.org/abs/1708.02759}{{\tt
  1708.02759}}].

\bibitem{Wondrak:2020tzt}
M.~F. Wondrak, M.~Kaminski and M.~Bleicher, \emph{{Shear transport far from
  equilibrium via holography}},
  \href{http://dx.doi.org/10.1016/j.physletb.2020.135973}{\emph{Phys. Lett. B}
  {\bf 811} (2020) 135973}, [\href{http://arxiv.org/abs/2002.11730}{{\tt
  2002.11730}}].

\bibitem{Baggioli:2021tzr}
M.~Baggioli, L.~Li and H.-T. Sun, \emph{{Shear Flows in Far-from-Equilibrium
  Strongly Coupled Fluids}},
  \href{http://dx.doi.org/10.1103/PhysRevLett.129.011602}{\emph{Phys. Rev.
  Lett.} {\bf 129} (2022) 011602}, [\href{http://arxiv.org/abs/2112.14855}{{\tt
  2112.14855}}].

\bibitem{huang2023revealing}
D.~Huang, M.~Baggioli, S.~Lu, Z.~Ma and Y.~Feng, \emph{Revealing the
  supercritical dynamics of dusty plasmas and their liquidlike to gaslike
  dynamical crossover},
  \href{http://dx.doi.org/10.1103/PhysRevResearch.5.013149}{\emph{Phys. Rev.
  Res.} {\bf 5} (Feb, 2023) 013149}.

\bibitem{PhysRevResearch.4.033064}
D.~Huang, S.~Lu, M.~S. Murillo and Y.~Feng, \emph{Origin of viscosity at
  individual particle level in yukawa liquids},
  \href{http://dx.doi.org/10.1103/PhysRevResearch.4.033064}{\emph{Phys. Rev.
  Res.} {\bf 4} (Jul, 2022) 033064}.

\bibitem{Cao:2011fh}
C.~Cao, E.~Elliott, H.~Wu and J.~E. Thomas, \emph{{Searching for Perfect
  Fluids: Quantum Viscosity in a Universal Fermi Gas}},
  \href{http://dx.doi.org/10.1088/1367-2630/13/7/075007}{\emph{New J. Phys.}
  {\bf 13} (2011) 075007}, [\href{http://arxiv.org/abs/1105.2496}{{\tt
  1105.2496}}].

\bibitem{PhysRevD.76.101701}
H.~B. Meyer, \emph{Calculation of the shear viscosity in su(3) gluodynamics},
  \href{http://dx.doi.org/10.1103/PhysRevD.76.101701}{\emph{Phys. Rev. D} {\bf
  76} (Nov, 2007) 101701}.

\bibitem{Adams:2012th}
A.~Adams, L.~D. Carr, T.~Sch\"afer, P.~Steinberg and J.~E. Thomas,
  \emph{{Strongly Correlated Quantum Fluids: Ultracold Quantum Gases, Quantum
  Chromodynamic Plasmas, and Holographic Duality}},
  \href{http://dx.doi.org/10.1088/1367-2630/14/11/115009}{\emph{New J. Phys.}
  {\bf 14} (2012) 115009}, [\href{http://arxiv.org/abs/1205.5180}{{\tt
  1205.5180}}].

\bibitem{Cremonini:2012ny}
S.~Cremonini, U.~Gursoy and P.~Szepietowski, \emph{{On the Temperature
  Dependence of the Shear Viscosity and Holography}},
  \href{http://dx.doi.org/10.1007/JHEP08(2012)167}{\emph{JHEP} {\bf 08} (2012)
  167}, [\href{http://arxiv.org/abs/1206.3581}{{\tt 1206.3581}}].

\bibitem{Cremonini:2011ej}
S.~Cremonini and P.~Szepietowski, \emph{{Generating Temperature Flow for eta/s
  with Higher Derivatives: From Lifshitz to AdS}},
  \href{http://dx.doi.org/10.1007/JHEP02(2012)038}{\emph{JHEP} {\bf 02} (2012)
  038}, [\href{http://arxiv.org/abs/1111.5623}{{\tt 1111.5623}}].

\bibitem{Gubser:2008wv}
S.~S. Gubser and S.~S. Pufu, \emph{{The Gravity dual of a p-wave
  superconductor}},
  \href{http://dx.doi.org/10.1088/1126-6708/2008/11/033}{\emph{JHEP} {\bf 11}
  (2008) 033}, [\href{http://arxiv.org/abs/0805.2960}{{\tt 0805.2960}}].

\bibitem{Cai:2015cya}
R.-G. Cai, L.~Li, L.-F. Li and R.-Q. Yang, \emph{{Introduction to Holographic
  Superconductor Models}},
  \href{http://dx.doi.org/10.1007/s11433-015-5676-5}{\emph{Sci. China Phys.
  Mech. Astron.} {\bf 58} (2015) 060401},
  [\href{http://arxiv.org/abs/1502.00437}{{\tt 1502.00437}}].

\bibitem{refId0}
{Parodi, O.}, \emph{Stress tensor for a nematic liquid crystal},
  \href{http://dx.doi.org/10.1051/jphys:01970003107058100}{\emph{J. Phys.
  France} {\bf 31} (1970) 581--584}.

\bibitem{doi:10.1146/annurev.fl.10.010178.001213}
J.~T. Jenkins, \emph{Flows of nematic liquid crystals},
  \href{http://dx.doi.org/10.1146/annurev.fl.10.010178.001213}{\emph{Annual
  Review of Fluid Mechanics} {\bf 10} (1978) 197--219}.

\bibitem{doi:10.1080/00268948108076134}
B.~C. Benicewicz, J.~F. Johnson and M.~T. Shaw, \emph{Viscosity behavior of
  liquid crystals},
  \href{http://dx.doi.org/10.1080/00268948108076134}{\emph{Molecular Crystals
  and Liquid Crystals} {\bf 65} (1981) 111--131}.

\bibitem{MIESOWICZ1946}
M.~Miesowicz, \emph{The three coefficients of viscosity of anisotropic
  liquids}, \href{http://dx.doi.org/10.1038/158027b0}{\emph{Nature} {\bf 158}
  (Jul, 1946) 27--27}.

\bibitem{doi:10.1063/1.465570}
S.~Sarman and D.~J. Evans, \emph{Statistical mechanics of viscous flow in
  nematic fluids}, \href{http://dx.doi.org/10.1063/1.465570}{\emph{The Journal
  of Chemical Physics} {\bf 99} (1993) 9021--9036}.

\bibitem{Erdmenger:2010xm}
J.~Erdmenger, P.~Kerner and H.~Zeller, \emph{{Non-universal shear viscosity
  from Einstein gravity}},
  \href{http://dx.doi.org/10.1016/j.physletb.2011.04.009}{\emph{Phys. Lett. B}
  {\bf 699} (2011) 301--304}, [\href{http://arxiv.org/abs/1011.5912}{{\tt
  1011.5912}}].

\bibitem{Erdmenger:2012zu}
J.~Erdmenger, D.~Fernandez and H.~Zeller, \emph{{New Transport Properties of
  Anisotropic Holographic Superfluids}},
  \href{http://dx.doi.org/10.1007/JHEP04(2013)049}{\emph{JHEP} {\bf 04} (2013)
  049}, [\href{http://arxiv.org/abs/1212.4838}{{\tt 1212.4838}}].

\bibitem{Grossi:2021gqi}
E.~Grossi, A.~Soloviev, D.~Teaney and F.~Yan, \emph{{Soft pions and transport
  near the chiral critical point}},
  \href{http://dx.doi.org/10.1103/PhysRevD.104.034025}{\emph{Phys. Rev. D} {\bf
  104} (2021) 034025}, [\href{http://arxiv.org/abs/2101.10847}{{\tt
  2101.10847}}].

\bibitem{Cao:2022csq}
X.~Cao, M.~Baggioli, H.~Liu and D.~Li, \emph{{Pion dynamics in a soft-wall
  AdS-QCD model}}, \href{http://dx.doi.org/10.1007/JHEP12(2022)113}{\emph{JHEP}
  {\bf 12} (2022) 113}, [\href{http://arxiv.org/abs/2210.09088}{{\tt
  2210.09088}}].

\bibitem{Ammon:2021slb}
M.~Ammon, D.~Arean, M.~Baggioli, S.~Gray and S.~Grieninger,
  \emph{{Pseudo-spontaneous $U(1)$ Symmetry Breaking in Hydrodynamics and
  Holography}},  \href{http://arxiv.org/abs/2111.10305}{{\tt 2111.10305}}.

\bibitem{Delacretaz:2021qqu}
L.~V. Delacr\'etaz, B.~Gout\'eraux and V.~Ziogas, \emph{{Damping of
  Pseudo-Goldstone Fields}},
  \href{http://dx.doi.org/10.1103/PhysRevLett.128.141601}{\emph{Phys. Rev.
  Lett.} {\bf 128} (2022) 141601}, [\href{http://arxiv.org/abs/2111.13459}{{\tt
  2111.13459}}].

\bibitem{Armas:2021vku}
J.~Armas, A.~Jain and R.~Lier, \emph{{Approximate symmetries,
  pseudo-Goldstones, and the second law of thermodynamics}},
  \href{http://arxiv.org/abs/2112.14373}{{\tt 2112.14373}}.

\bibitem{JanJadzyn_2001}
J.~Jadzyn and G.~Czechowski, \emph{The shear viscosity minimum of freely
  flowing nematic liquid crystals},
  \href{http://dx.doi.org/10.1088/0953-8984/13/12/101}{\emph{Journal of
  Physics: Condensed Matter} {\bf 13} (mar, 2001) L261}.

\bibitem{diogo:jpa-00209451}
A.~Diogo and A.~Martins, \emph{{Order parameter and temperature dependence of
  the hydrodynamic viscosities of nematic liquid crystals}},
  \href{http://dx.doi.org/10.1051/jphys:01982004305077900}{\emph{{Journal de
  Physique}} {\bf 43} (1982) 779--786}.

\bibitem{Chen:15}
H.~Chen, M.~Hu, F.~Peng, J.~Li, Z.~An and S.-T. Wu, \emph{Ultra-low viscosity
  liquid crystal materials},
  \href{http://dx.doi.org/10.1364/OME.5.000655}{\emph{Opt. Mater. Express} {\bf
  5} (Mar, 2015) 655--660}.

\bibitem{D1CC06111A}
R.~Kimura, H.~Kitakado, T.~Yamakado, H.~Yoshida and S.~Saito, \emph{Probing a
  microviscosity change at the nematic–isotropic liquid crystal phase
  transition by a ratiometric flapping fluorophore},
  \href{http://dx.doi.org/10.1039/D1CC06111A}{\emph{Chem. Commun.} {\bf 58}
  (2022) 2128--2131}.

\bibitem{Cai:2021obq}
R.-G. Cai, C.~Ge, L.~Li and R.-Q. Yang, \emph{{Inside anisotropic black hole
  with vector hair}},
  \href{http://dx.doi.org/10.1007/JHEP02(2022)139}{\emph{JHEP} {\bf 02} (2022)
  139}, [\href{http://arxiv.org/abs/2112.04206}{{\tt 2112.04206}}].

\bibitem{Natsuume:2014sfa}
M.~Natsuume, \emph{{AdS/CFT Duality User Guide}}, vol.~903.
\newblock 2015,
  \href{http://dx.doi.org/10.1007/978-4-431-55441-7}{10.1007/978-4-431-55441-7}.

\bibitem{Baggioli:2018bfa}
M.~Baggioli and A.~Buchel, \emph{{Holographic Viscoelastic Hydrodynamics}},
  \href{http://dx.doi.org/10.1007/JHEP03(2019)146}{\emph{JHEP} {\bf 03} (2019)
  146}, [\href{http://arxiv.org/abs/1805.06756}{{\tt 1805.06756}}].

\bibitem{Ammon:2021pyz}
M.~Ammon, D.~Arean, M.~Baggioli, S.~Gray and S.~Grieninger,
  \emph{{Pseudo-spontaneous $U(1)$ symmetry breaking in hydrodynamics and
  holography}}, \href{http://dx.doi.org/10.1007/JHEP03(2022)015}{\emph{JHEP}
  {\bf 03} (2022) 015}, [\href{http://arxiv.org/abs/2111.10305}{{\tt
  2111.10305}}].

\bibitem{Amado:2009ts}
I.~Amado, M.~Kaminski and K.~Landsteiner, \emph{{Hydrodynamics of Holographic
  Superconductors}},
  \href{http://dx.doi.org/10.1088/1126-6708/2009/05/021}{\emph{JHEP} {\bf 05}
  (2009) 021}, [\href{http://arxiv.org/abs/0903.2209}{{\tt 0903.2209}}].

\bibitem{Arean:2021tks}
D.~Arean, M.~Baggioli, S.~Grieninger and K.~Landsteiner, \emph{{A holographic
  superfluid symphony}},
  \href{http://dx.doi.org/10.1007/JHEP11(2021)206}{\emph{JHEP} {\bf 11} (2021)
  206}, [\href{http://arxiv.org/abs/2107.08802}{{\tt 2107.08802}}].

\bibitem{Cremonini:2014pca}
S.~Cremonini, X.~Dong, J.~Rong and K.~Sun, \emph{{Holographic RG flows with
  nematic IR phases}},
  \href{http://dx.doi.org/10.1007/JHEP07(2015)082}{\emph{JHEP} {\bf 07} (2015)
  082}, [\href{http://arxiv.org/abs/1412.8638}{{\tt 1412.8638}}].

\bibitem{Cai:2013aca}
R.-G. Cai, L.~Li and L.-F. Li, \emph{{A Holographic P-wave Superconductor
  Model}}, \href{http://dx.doi.org/10.1007/JHEP01(2014)032}{\emph{JHEP} {\bf
  01} (2014) 032}, [\href{http://arxiv.org/abs/1309.4877}{{\tt 1309.4877}}].

\bibitem{Garbayo:2022pqp}
A.~Garbayo, C.~Hoyos, N.~Jokela, J.~M. Pen\'\i{}n and A.~V. Ramallo,
  \emph{{Flavored anisotropic black holes}},
  \href{http://dx.doi.org/10.1007/JHEP10(2022)061}{\emph{JHEP} {\bf 10} (2022)
  061}, [\href{http://arxiv.org/abs/2208.04958}{{\tt 2208.04958}}].

\bibitem{Hoyos:2020zeg}
C.~Hoyos, N.~Jokela, J.~M. Pen\'\i{}n and A.~V. Ramallo, \emph{{Holographic
  spontaneous anisotropy}},
  \href{http://dx.doi.org/10.1007/JHEP04(2020)062}{\emph{JHEP} {\bf 04} (2020)
  062}, [\href{http://arxiv.org/abs/2001.08218}{{\tt 2001.08218}}].

\bibitem{Ji:2022ovs}
T.~Ji, L.~Li and H.-T. Sun, \emph{{Thermoelectric transport in holographic
  quantum matter under shear strain}},
  \href{http://dx.doi.org/10.1088/1572-9494/aca0e1}{\emph{Commun. Theor. Phys.}
  {\bf 75} (2023) 015401}, [\href{http://arxiv.org/abs/2208.08803}{{\tt
  2208.08803}}].

\bibitem{Baggioli:2020qdg}
M.~Baggioli, V.~C. Castillo and O.~Pujolas, \emph{{Black Rubber and the
  Non-linear Elastic Response of Scale Invariant Solids}},
  \href{http://dx.doi.org/10.1007/JHEP09(2020)013}{\emph{JHEP} {\bf 09} (2020)
  013}, [\href{http://arxiv.org/abs/2006.10774}{{\tt 2006.10774}}].

\end{thebibliography}\endgroup

\end{document}